\pdfoutput=1
\documentclass[twocolumn,nofootinbib,9pt]{revtex4-1}
\usepackage{amsmath,amssymb}    % need for subequations
\usepackage[dvipdfmx]{graphicx} % 大概これで解決
\usepackage{color}
\usepackage{subfigure}
\usepackage{here}
\usepackage{type1cm}%文字サイズの自由度を広げる
\usepackage{bm}
\usepackage{tabularx}
\usepackage{enumerate}
\usepackage{here}
\usepackage{comment}%コメントアウト
\usepackage{afterpage}%改ページできるようになる
\usepackage{ascmac}%囲い枠が使えるようになる
\usepackage{listings}%ソースコードが書けるようになる
\begin{document}

\title{
Flexible Two-point Selection Approach for Characteristic Function-based Parameter Estimation of Stable Laws
%Characteristic Function-based Parameter Estimation of Stable Laws with Appropriate Point Selection
%An Analytical Approach Based on Empirical Characteristic Function with Sensitive Selection for Parameter Estimation of Stable Laws
}
\author{Shinji Kakinaka and Ken Umeno}
\affiliation{Department of Applied Mathematics and Physics,\\
Graduate School of Informatics, Kyoto University, Japan.}

\date{\today}

\begin{abstract}
Stable distribution is one of the attractive models that well describes fat-tail behaviors and scaling phenomena in various scientific fields. The approach based upon the method of moments yields a simple procedure for estimating stable law parameters with the requirement of using momental points for the characteristic function, but the selection of points is only poorly explained and has not been elaborated. We propose a new characteristic function-based approach by introducing a technique of selecting plausible points, which could bring the method of moments available for practical use. Our method outperforms other state-of-art methods that exhibit a closed-form expression of all four parameters of stable laws. Finally, the applicability of the method is illustrated by using several data of financial assets. Numerical results reveal that our approach is advantageous when modeling empirical data with stable distributions.
\end{abstract}
\maketitle

\section{Introduction}
A fundamental theory of stochastic processes in various scientific fields is the generalized central limit theorem (GCLT), which points out that the sum of independent and identically distributed random variables converges only to the family of stable distribution~\cite{Gnedenko1954}.
There are some challenges to overcome the analytic difficulties of stable distributions since the probability density function (PDF) is not always expressed in a closed form.
Numerically approximated expressions are known in symmetric cases ($\beta=0$) based on hypergeometric functions, but those in unrestricted asymmetric cases are often too complex for estimating the parameters of the stable distribution~\cite{Crisanto2018}.
More practically, the estimation of all parameters is the most basic and necessary process for any application, but it remains to be one of the most controversial issues when attempting to detect stable laws.
Numerous approaches have been studied for the parameter estimation.
The primary approaches include the approximate maximum likelihood estimation~\cite{DuMouchel1973, BrorsenYang1990, Mittnik1999, Nolan2001}, the bayesian based method~\cite{Koblents2016}, the quantile method (QM)~\cite{FamaRoll1971, McCulloch1986}, the fractional lower order moment (FLOM) method~\cite{Ma1995,Kuruoglu2001}, the method of log-cumulant (MLOC)~\cite{Nicolas2002, Pastor2016}, the characteristic function-based (CF-based) method~\cite{Koutrouvelis1980, Press1972, Bibalan2017, Krutto2016, Krutto2018}, and their hybrid combinations.
Many of them tend to have different kinds of drawbacks, such as restrictions of parameter ranges, complex estimation algorithms, high computational costs, requirements of larger datasets, and low accuracy.
To the best of our knowledge, the FLOM, MOLC, and QM and some class of the CF-based methods~\cite{Press1972, Bibalan2017, Krutto2016, Krutto2018} provide closed-form estimators of stable laws.

The CF-based method is perhaps the largest classification group, including a variety of methods and approaches developed under different techniques.
In particluar, Press (1972)~\cite{Press1972} presents the {\it method of moments}, which offers a simple approach to estimate all four parameters of stable distribution using the characteristic function (CF) evaluated at four arbitrary points.
The biggest advantage of this method is that it is likely to have less drawbacks compared to other primary methods, but it carries a fundamental problem.
Without appropriate points given, the performance is poor, and unfortunately Press leaves unsolved the crucial idea about the choice of points at which the CF should be evaluated.
The selection of the points has long been an open question, although several studies have made an effort to improve the method of moments by reducing the use of points from four to two and discussing their choice.
Krutto (2016, 2018) provides some guidance on how the two positive points should be chosen through empirical searches relying on the cumulant function~\cite{Krutto2016, Krutto2018}.
Bibalan et al. (2017) focus on the absolute value of the CF and suggest an algorithmic approach where a positive point is fixed for each scaling parameter~\cite{Bibalan2017}.
They show accurate estimates within certain parameter ranges, but their method fails to support a wider range of parameter spaces.
Thus, these approaches are not comprehensive, so that the method of detecting more appropriate points related to the CF is required for practical uses.

In this paper, we propose an effective and practical method for estimating stable laws.
We greatly improve the method of moments by introducing a new technique for the selection of two positive points at which the CF is evaluated.
The technique is developed over the extension of both algorithmic and empirical search approaches.
The idea of empirical search plays a role in determining the scaling related estimates, which take crucial responsibility for indicating statistical values derived in the estimation process, whereas the concept of the algorithmic approach yields various ideas of inferences based on the absolute value of the CF.
Our approach realizes the possibility of choosing different values of points depending on the index parameter $\alpha$, which is a new perspective.
We assess and compare the performance of our method to those of other methods in terms of the Mean Squared Error (MSE) criterion and the Kolmogorov-Smirnov (KS) distance.
Our proposed method generally outperforms all the other state-of-art methods that exhibit closed-form expressions for all four parameters of stable laws.
It is practically straightforward and assures that there is no restriction of parameter ranges, except for $\alpha=1$ due to the discontinuous form of the one-parameterization CF.
Finally, we apply our method to price fluctuation behaviors of several financial assets to examine the appropriateness for practical uses.

This paper is organized as follows.
Section 2 shows preliminaries on stable distribution and its basic properties.
We follow in the next section to describe the existing methods for estimating the parameters of stable laws.
In section 4 we propose a new technique of the CF-based parameter estimation method.
The arguments for the selection of points at which the CF should be evaluated are discussed.
In section 5 we report the performance with the comparison to other representative methods and present that our method provides accurate estimates of stable distribution.
The last section shows application to financial data and confirms that our method is applicable for empirical studies.
%==============================================================================
%%%%%%%%%%%%%%%%%%%%%%%%%%%%%%%%%%%%%%%%%%%%%%%%%%%%
% 2. Stable distribution
%%%%%%%%%%%%%%%%%%%%%%%%%%%%%%%%%%%%%%%%%%%%%%%%%%%%
\section{Stable Distribution}
In this section, we summarize the basis and properties of the stable distribution.
We explain the definition of the stable distribution and its properties.
%%%%%%%%%%%%%%%%%%%%%%%%%%%%%%%%%%%%%%%%%%%%%%%%%%%%
% 2.1 Basis of stable distribution
%%%%%%%%%%%%%%%%%%%%%%%%%%%%%%%%%%%%%%%%%%%%%%%%%%%%
\subsection{Basis of stable distribution}
Stable distribution, also known as $\alpha$-stable distribution, or L\'evy's stable distribution, was first introduced by Paul L\'evy (1937)~\cite{PaulLevy1937}, which is a family of parametric distribution with tails that are expressed as power-functions.
In the far tails the PDF can be written as~\cite{SamoTaqqu1994a},
\begin{align*}
	%\label{eq:stable_fartailPDF}
	f(x;\alpha,\beta,\gamma,\delta) \simeq
	\begin{cases}
	c_\alpha \gamma^{\alpha} \alpha\,(1+\beta)\,|x|^{-(1+\alpha)} \;\; \mbox{for} \;\; (x \rightarrow +\infty) \\
	c_\alpha \gamma^{\alpha} \alpha\,(1-\beta)\,|x|^{-(1+\alpha)} \;\; \mbox{for} \;\; (x \rightarrow -\infty),
	\end{cases}
\end{align*}
and the cumulative distribution function (CDF) written as,
\begin{align*}
	%\label{eq:stable_fartailCDF}
	\begin{cases}
	P(X>x) \simeq
	c_\alpha \gamma^{\alpha}(1+\beta)\,|x|^{-\alpha} \;\; \mbox{for} \;\; (x \rightarrow +\infty) \\
	P(X<x) \simeq
	c_\alpha \gamma^{\alpha}(1-\beta)\,|x|^{-\alpha} \;\; \mbox{for} \;\; (x \rightarrow -\infty),
	\end{cases}
\end{align*}
where $c_\alpha$ is a constant value $[\sin (\pi \alpha / 2)\Gamma(\alpha)]/\pi$.
Stable distribution is represented by four parameters; the tail index parameter $\alpha \in (0,2]$ representing the fatness of the tail, the skewness parameter $\beta \in [-1,1]$, the scaling parameter $\gamma >0$, and the location parameter $\delta \in \mathbb{R}$.
Especially the parameters $\alpha$ and $\beta$ determine the shape of distribution, including various forms of widely-known distributions such as the Gaussian and Cauchy distribution.
Smaller value of $\alpha$ indicates fatter tails and hence it is well known that the variance diverges for ${0 < \alpha < 2}$, and also the mean cannot be defined for ${0 < \alpha \leq 1}$.
Note that if $\beta=0$, the distribution is symmetric, if $\beta>0$, right-tailed, and if $\beta<0$, left-tailed.

The definition of stable distribution is that the linear combination of independent random variables that follow a stable distribution with tail index $\alpha$ invariably becomes again a stable distribution with the same tail index $\alpha$.
More particularly, when variables $X_1, X_2$ are i.i.d. copies of a random variable $X$ and ${a,b}$ are positive constant numbers, $X$ is said to be {\it stable} and follows a stable distribution if there is a positive constant number $c$ and a real number $d\in \mathbb{R}$ that satisfies
\begin{align*}
	aX_1+bX_2 \overset{\mathrm{d}}{=} cX+d,
\end{align*}
also known for {\it stability property}.
When a variable $X$ follows a stable distribution, the notation $X \overset{\mathrm{d}}{=} S(\alpha, \beta, \gamma, \delta)$ is often used, where ${\overset{\mathrm{d}}{=}}$ denotes equality in distribution~\cite{SamoTaqqu1994}.
Variable $X$ can be standardized according to the following property:
\begin{align}
	\label{eq:standardize}
	\frac{X-\delta}{\gamma} \overset{\mathrm{d}}{=} S(\alpha, \beta, 0, 1).
\end{align}

Another important property of stable distribution is the GCLT, which implies that the only possible limit distributions for sums of i.i.d random variables is a family of stable distribution.
When $\alpha = 2$, that is, when i.i.d. random variables have finite variance, the limit distribution then becomes a Gaussian according to the well-known classical Central Limit Theorem (CLT).
%%%%%%%%%%%%%%%%%%%%%%%%%%%%%%%%%%%%%%%%%%%%%%%%%%%%
% 2.2 Characteristic function
%%%%%%%%%%%%%%%%%%%%%%%%%%%%%%%%%%%%%%%%%%%%%%%%%%%%
\subsection{Characteristic function}
\begin{figure}
 \centering
  \subfigure[stable distribution for the case of $S(\alpha,0,1,0)$]{\includegraphics[width=\linewidth]{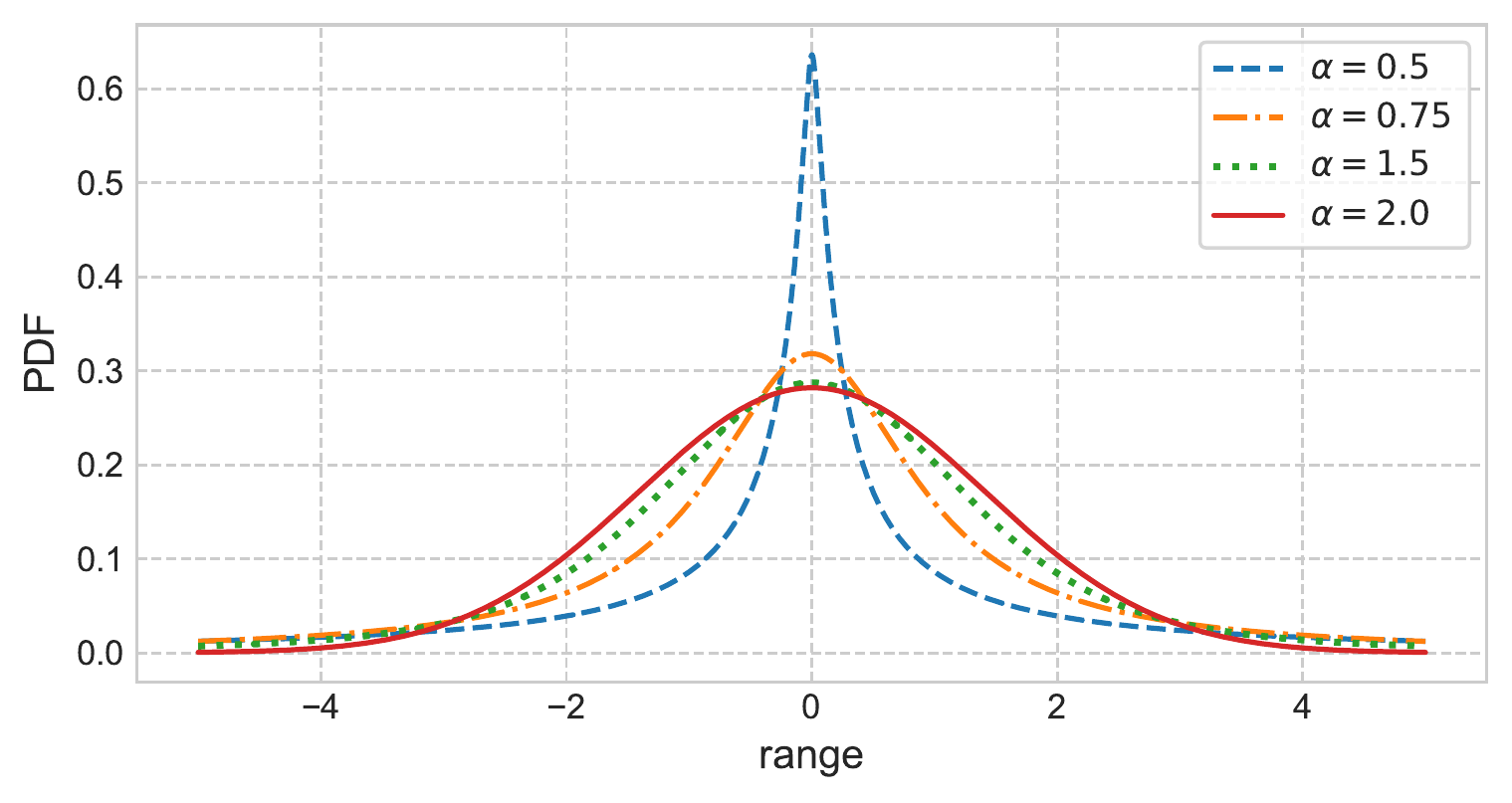}}\\
  \subfigure[stable distribution for the case of $S(0.5,\beta,1,0)$]{\includegraphics[width=\linewidth]{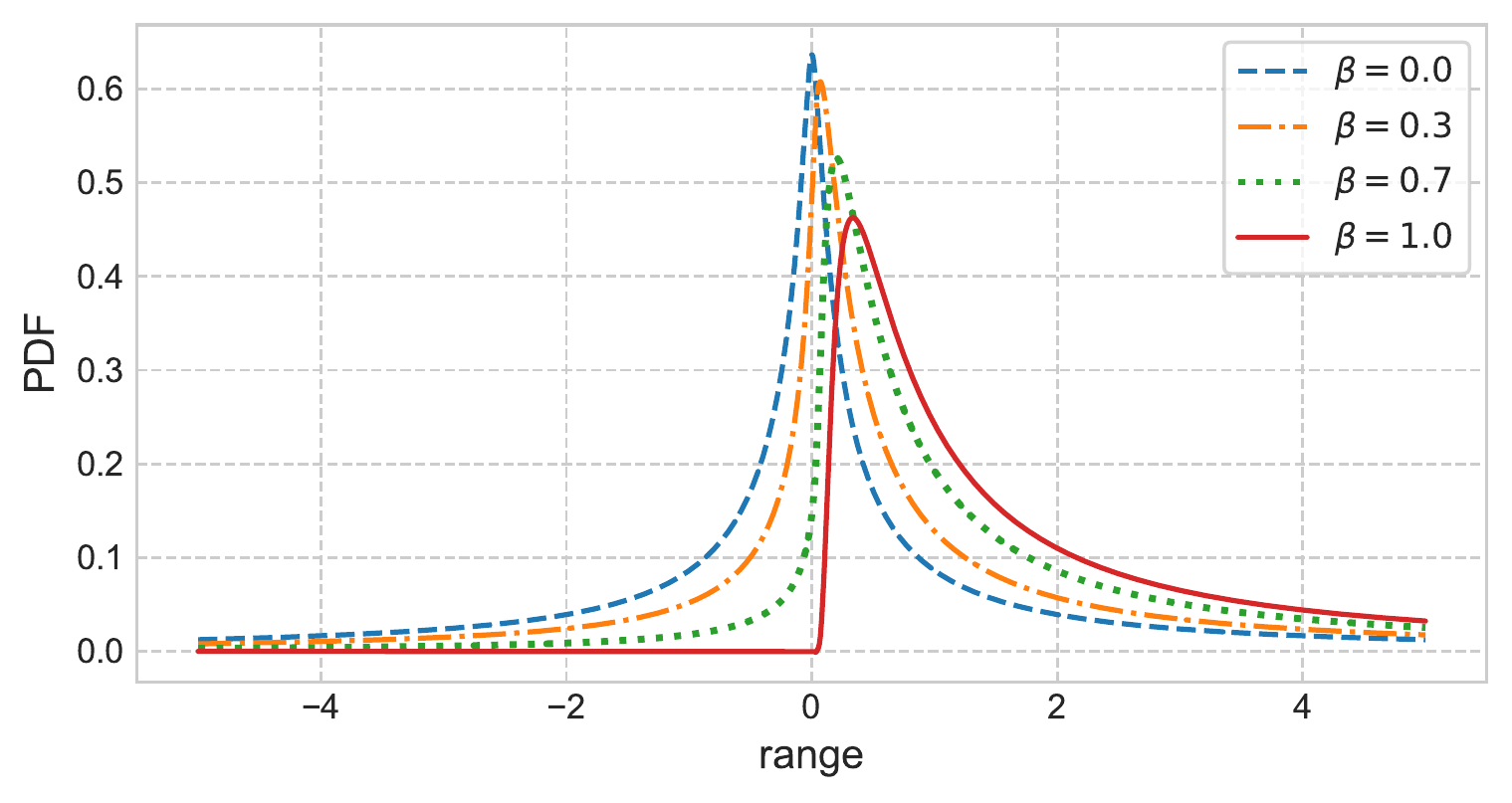}}
  \caption{
  Standardized stable distributions with the one-parameterization form for different parameters of $\alpha$ and $\beta$.
   } 
   \label{stable_distribution}
\end{figure}

The PDF of stable distribution cannot be written in a closed form except for some cases; Cauchy distribution (${\alpha=1,\,\beta=0}$), L\'evy distribution (${\alpha=1/2,\,\beta=1}$), and Gaussian distribution (${\alpha=2}$).
Alternatively, the features are expressed by the characteristic function (CF), ${\varphi(k)}$, which is the Fourier transform of the PDF.
By taking the inverse Fourier transform of the CF, the PDF can be obtained as
\begin{align*}
	%\label{eq:FT}
	f(x)=\frac{1}{2\pi} \int_{-\infty}^{\infty} \mathrm{e}^{-ikx} \varphi(k) dk.
\end{align*}
When variable $X$ follows a stable distribution with ${S(\alpha,\beta,\gamma,\delta)}$, the CF is shown as
\begin{align}
	\label{eq:first}
	\nonumber
	\varphi(k) &=
	\exp \left \{ i\delta k -\gamma^\alpha |k|^\alpha \left (1-i \beta \operatorname{sgn}(k) \omega(k, \alpha)\right) \right \}, \\
	\omega(k, \alpha) &=
	\begin{cases}
	\tan( \frac{\pi \alpha}{2} ) & \alpha \neq 1 \\
	-\frac{2}{\pi} \log|k| & \alpha = 1,
	\end{cases}
\end{align}
which corresponds to the one-parameterization form of $S(\alpha,\beta,\gamma,\delta;1)$ in Nolan (2003)~\cite{Nolan2018}.
This is the most popular parameterization among many other forms of the stable distribution owing to the simplicity of the form.
Figure~\ref{stable_distribution} shows the standardized stable distributions with the one-parameterization form for different parameters of $\alpha$ and $\beta$, as an example.

One-parameterization is preferred when one is interested in the basic properties of the distribution, but the CF takes a discontinuous form at $\alpha=1$.
Nolan suggests the use of the zero-parameterization form $S(\alpha,\beta,\gamma,\delta_0;0)$ with different $\omega(k,\alpha)$ shown as
\begin{align}
	\label{eq:first2}
	\omega(k, \alpha) &=
	\begin{cases}
	-\left (|\gamma k|^{1-\alpha}-1 \right) \tan( \frac{\pi \alpha}{2} ) & \alpha \neq 1 \\
	-\frac{2}{\pi} \log|\gamma k| & \alpha = 1,
	\end{cases}
\end{align}
giving a more complex form, but provides a continuous form.
The only difference between the parameterization is the location parameter, which they are related by
\begin{align}
	\label{eq:parameterization_relation}
	\nonumber
    \delta_{0}=\left\{\begin{array}{ll}{\delta+\beta \gamma \tan \frac{\pi \alpha}{2}} & {\alpha \neq 1} \\ {\delta+\beta \frac{2}{\pi} \gamma \log \gamma} & {\alpha=1}\end{array},\right.\\
    \delta=\left\{\begin{array}{ll}{\delta_{0}-\beta \gamma \tan \frac{\pi \alpha}{2}} & {\alpha \neq 1} \\ {\delta_{0}-\beta \frac{2}{\pi} \gamma \log \gamma} & {\alpha=1}\end{array}.\right.
\end{align}

In this paper, we employ the simple one-parameterization, as we are interested in estimating the four parameters through the CF, and many existing estimation methods comply with that form.
However, since this CF does not have a continuous form at $\alpha=1$, arguments with different parameterizations may be more appropriate for discussing distributions when we already know that $\alpha$ is 1, for instance, the case of Cauchy distribution ($\alpha=1, \beta=0$).
%==============================================================================
%%%%%%%%%%%%%%%%%%%%%%%%%%%%%%%%%%%%%%%%%%%%%%%%%%%%
% 3. Parameter Estimation of Stable Laws
%%%%%%%%%%%%%%%%%%%%%%%%%%%%%%%%%%%%%%%%%%%%%%%%%%%%
\section{Parameter Estimation of Stable Laws}
This section gives an overview of the methods for the parameter estimation of the stable distribution.
We review two major methods, both of which are considered as an analytical approach that provides a closed-form expression of the estimates--- the quantile method and the characteristic function-based method (CF-based method).
Several different approaches are explained for the CF-based method.
%%%%%%%%%%%%%%%%%%%%%%%%%%%%%%%%%%%%%%%%%%%%%%%%%%%%
% 3.1 Quantile method
%%%%%%%%%%%%%%%%%%%%%%%%%%%%%%%%%%%%%%%%%%%%%%%%%%%% 
\subsection{Quantile method}
McCulloch (1986) proposes the use of five sample quantiles $x_{0.05},x_{0.25},x_{0.5},x_{0.75}$, and $x_{0.95}$ as an informative measure for estimating the four parameters of stable laws, known as the quantile method (QM)~\cite{McCulloch1986}.
He improves the former method of Fama \& Roll (1971) by eliminating bias in estimates and relaxing estimation restrictions~\cite{FamaRoll1971}.
The idea is to calculate the functions $\phi_i(\alpha, \beta) \:(i=1,2,3,4)$, where the relationships between the function values and the parameters are already studied and known beforehand.
The method first sets out to estimate $\alpha$ and $\beta$ by using the functions $\phi_1(\alpha, \beta)$ and $\phi_2(\alpha, \beta)$ independent of both $\gamma$ and $\delta$ defined as
\begin{align}
	\label{quantile1}
	\phi_{1}(\alpha, \beta)&=\frac{x_{0.95}-x_{0.05}}{x_{0.75}-x_{0.25}}\\
	\label{quantile2}
	\phi_{2}(\alpha, \beta)&=\frac{\left(x_{0.95}-x_{0.5}\right)-\left(x_{0.5}-x_{0.05}\right)}{x_{0.95}-x_{0.05}}.
\end{align}
Equation~\eqref{quantile1} refers to the measure of fat-tail behaviors with the focus on estimating $\alpha$, and equation~\eqref{quantile2} is a measure of skewness effects with the focus on estimating $\beta$.
With empirical values of sample quantiles and employing linear interpolation with tabular look-ups, the estimates $\hat{\alpha}, \hat{\beta}$ are inversely obtained. 
To avoid $\hat{\alpha}$ being larger than 2, outside the parameter range, $\hat{\phi}_1 = \frac{\left(\hat{x}_{0.95}-\hat{x}_{0.5}\right)-\left(\hat{x}_{0.5}-\hat{x}_{0.05}\right)}{\hat{x}_{0.95}-\hat{x}_{0.05}}$ can be no larger than the upper range 2.439, which corresponds to the case of $\alpha=2$ (note that $\beta$ is not identified in this case).

Next, the scale and location parameter $\gamma$ and $\delta$ can be estimated using the functions defined as
\begin{align}
	\label{quantile3}
	\phi_{3}(\alpha, \beta)&=\frac{x_{0.75}-x_{0.25}}{\gamma}\\
	\label{quantile4}
	\phi_{4}(\alpha, \beta)&=\frac{\mu-x_{0.5}}{\gamma}+\beta \tan \left(\frac{\pi \alpha}{2} \right).
\end{align}
The function $\phi_{3}(\alpha, \beta)$ indicates the standardized form of sample sizes for the middle part of distribution.
Since it does not depend on $\gamma$ nor $\delta$, the value can be informed by tabular look-ups based on $\alpha$ and $\beta$, which the relations are studied and known beforehand.
After calculating $\hat{\gamma} = \frac{\hat{x}_{0.75}-\hat{x}_{0.25}}{\hat{\phi}_{3}(\hat{\alpha},\,\hat{\beta})}$ in equation~\eqref{quantile3}, the location parameter $\delta$ can be estimated from equation~\eqref{quantile4} using the values $\hat{\phi}_{4}(\hat{\alpha}, \hat{\beta})$ and $\hat{\gamma}$.
The relations of the parameter values and the function value $\phi_{4}(\alpha, \beta)$ are again, studied and known beforehand. 
In the case of $\alpha=1$, $\phi_{4}(\alpha, \beta)$ diverges and we cannot obtain the estimates for $\delta$.
McCulloch therefore suggests a complicated approach to overcome the discontinuity of the stable CF.
The method improves other issues and provides accurate estimates, however, it has parameter restrictions and can be applied only when $\alpha \geq 0.6$.
%%%%%%%%%%%%%%%%%%%%%%%%%%%%%%%%%%%%%%%%%%%%%%%%%%%%
% 3.2 Characteristic function-based method
%%%%%%%%%%%%%%%%%%%%%%%%%%%%%%%%%%%%%%%%%%%%%%%%%%%%
\subsection{Characteristic function-based method}
The CF-based method relies on the use of a consistent estimator of the CF $\varphi(k)$ for any fixed $k$.
The advantage of this method essentially lies in the fact that the stable CF can be expressed explicitly, making discussions straightforward compared to methods based on other distribution forms.
Under the assumption that given data $X_n\,(n=1,2,\ldots,N)$ are {\it ergodic}~\cite{ArnoldAvez}, the CF is obtained empirically by the following equation,
\begin{align}
	\label{eq_chaosfourier}
	\hat{\varphi} \left(k\right)=\frac{1}{N} \sum_{n=1}^{N} \mathrm{e}^{i k X_{n}}.
\end{align}

There are several approaches for estimating parameters of stable laws that take advantage of the explicit form of CF.
Koutrouvelis (1980)~\cite{Koutrouvelis1980} proposed a regression-type approach, which employs the iteration of two regression runs.
Moreover, the regression of the method requires different values of initial points $k$ depending on initial estimates of the parameters and sample sizes.
The number of points necessary for the regression also varies over initial conditions.
Although the accuracy of $\beta$ is unsatisfactory in some cases, the method generally shows accurate estimates of $\alpha$, and hence it is often suggested as a practical method for empirical analysis~\cite{Wang2015, Kateregga2017}.
However, some studies compare the method to McCulloch's quantile method and report that the regression-type method does not significantly improve the classical quantile method~\cite{Akgiray1989, Rene2011}, especially for $\alpha$ smaller than 1.
Other studies simplified the method by eliminating the iteration process and fixing the initial points to some extent, but still leaves behind the issues of estimating when $\alpha$ is small~\cite{Kogon1998, Borak2005}.
We do not consider the regression-type approach in this paper as the method generally relies on iteration and the estimates cannot be written analytically.

Another approach is based on the method of moment~\cite{Press1972}, which was later remodeled and simplified with the use of two given points of the CF~\cite{Krutto2016,Bibalan2017,Krutto2018}.
Starting off with the CF with the points $k_0$ and $k_1$, taking the absolute value cancels out the effect of parameters $\beta$ and $\delta$, and we obtain
\begin{align}
	\begin{cases}
		\label{eq_absCF1}
		|\varphi(k_0 ; \alpha, \beta, \gamma, \delta)| = \exp(-\gamma^{\alpha}|k_0|^{\alpha}) \\
		|\varphi(k_1 ; \alpha, \beta, \gamma, \delta)| = \exp(-\gamma^{\alpha}|k_1|^{\alpha}).
	\end{cases}
\end{align}
Taking the cumulant function, which is the natural logarithm of the CF, leads to the same discussion neutralizing the effect of parameters $\beta$ and $\delta$.
The equation $\ln \varphi=\ln |\varphi|+j(\arg \varphi+2 n \pi)$ implies that the real part of the cumulant function corresponds to the natural logarithm of the absolute value of CF, shown as
\begin{align}
	\begin{cases}
	\label{eq_CF_koutoshiki1}
	\Re \left \{ \ln \varphi(k_0; \alpha, \beta, \gamma, \delta) \right \} = \ln |\varphi(k_0 ; \alpha, \beta, \gamma, \delta)| = -\gamma^{\alpha}|k_0|^{\alpha} \\
	\Re \left \{ \ln \varphi(k_0; \alpha, \beta, \gamma, \delta) \right \} = \ln |\varphi(k_1 ; \alpha, \beta, \gamma, \delta)| = -\gamma^{\alpha}|k_1|^{\alpha},
	\end{cases}
\end{align}
for any value of $k$.
We consider only the positive values for convenience, since the CF is a symmetric function. 
By solving the above equations simultaneously, parameters $\alpha$ and $\gamma$ can be estimated shown as
\begin{align}
	\label{eq_alpha_estimate}
	&\hat{\alpha} = \frac{\ln \left ( -\Re \left \{ \ln \hat{\varphi} \left( k_0 \right) \right \} \right ) - \ln \left ( -\Re \left \{ \ln \hat{\varphi} \left( k_1 \right) \right \} \right ) }{\ln k_0 - \ln k_1}, \\ \nonumber
	&\hat{\gamma} =\\
	\label{eq_gamma_estimate}
	&\exp \left\{ \frac{\ln k_0 \ln \left ( -\Re \left \{ \ln \hat{\varphi} \left( k_1 \right) \right \} \right ) - \ln k_1 \ln \left ( -\Re \left \{ \ln \hat{\varphi} \left( k_0 \right) \right \} \right ) }{\ln \left( -\Re \left\{ \ln \hat{\varphi} \left( k_0 \right) \right\} \right) - \ln \left( -\Re \left\{ \ln \hat{\varphi} \left( k_1 \right) \right\} \right)} \right\}.
\end{align}

Since the one-parameterization form in equation~\eqref{eq:first} is discontinuous at $\alpha = 1$, the estimation of the remaining parameters $\beta$ and $\delta$ is divided into two cases.
When $\alpha \neq 1$, the cumulant function of stable distributions with the points $k_0, k_1 > 0$ are
\begin{align}
	\begin{cases}
		\label{eq_cumulantfunction1}
		\vspace{0.5em}
		\ln \varphi(k_0 ; \alpha, \beta, \gamma, \delta) = -\gamma^\alpha {k_0}^\alpha + i\left[\delta k_0 + \gamma^\alpha {k_0}^\alpha \beta \tan \left(\frac{\pi \alpha}{2}\right) \right] \\
		\ln \varphi(k_1 ; \alpha, \beta, \gamma, \delta) = -\gamma^\alpha {k_1}^\alpha + i\left[\delta k_1 + \gamma^\alpha {k_1}^\alpha \beta \tan \left(\frac{\pi \alpha}{2}\right) \right].
	\end{cases}
\end{align}
As we need the information of the parameters $\beta$ and $\delta$, we take the imaginary part.
Then the parameters $\beta$ and $\delta$ are estimated by solving the above equations simultaneously and using the estimates $\hat{\alpha}$ and $\hat{\gamma}$:
\begin{align}
	\label{eq_beta_estimate}
	\hat{\beta} &= \frac{k_1 \Im \left \{ \ln \hat{\varphi}(k_0) \right \} -k_0 \Im \left \{ \ln \hat{\varphi}(k_1) \right \} }{\hat{\gamma}^{\hat{\alpha}} \tan \left( \frac{\pi \hat{\alpha}}{2} \right) ({k_0}^{\hat{\alpha}} k_1 - {k_1}^{\hat{\alpha}}k_0)} \\
	\label{eq_delta_estimate}
	\hat{\delta} &= \frac{{k_1}^{\hat{\alpha}} \Im \left \{ \ln \hat{\varphi}(k_0) \right \} -{k_0}^{\hat{\alpha}} \Im \left \{ \ln \hat{\varphi}(k_1) \right \} }{k_0 {k_1}^{\hat{\alpha}} - k_1 {k_0}^{\hat{\alpha}}}.
\end{align}
In the case of $\alpha = 1$, the CF takes a discontinuous form and the cumulant functions are written as
\begin{align}
	\begin{cases}
		\vspace{0.5em}
		\label{eq_cumulantfunction3}
		\ln \varphi(k_0 ; 1, \beta, \gamma, \delta) = -\gamma {k_0} + i\left[\delta k_0 - \beta \frac{2}{\pi} \ln k_0 \right] \\
		\ln \varphi(k_1 ; 1, \beta, \gamma, \delta) = -\gamma {k_1} + i\left[\delta k_1 - \beta \frac{2}{\pi} \ln k_1 \right].
	\end{cases}
\end{align}
Then the parameters are estimated by solving the above equations simultaneously as well:
\begin{align}
	\label{eq_beta_estimate2}
	\hat{\beta} &= \frac{\pi}{2} \frac{k_1 \Im \left \{ \ln \hat{\varphi}(k_0) \right \} -k_0 \Im \left \{ \ln \hat{\varphi}(k_1) \right \} }{\hat{\gamma} k_0 k_1 \left( \ln k_1 - \ln k_0 \right)} \\
	\label{eq_delta_estimate2}
	\hat{\delta} &= \frac{{k_1} \Im \left \{ \ln \hat{\varphi}(k_0) \right \} \ln k_1 - {k_0} \Im \left \{ \ln \hat{\varphi}(k_1) \right \} \ln k_0 }{k_0 k_1 \left( \ln k_1 - \ln k_0 \right)}.
\end{align}
For simplicity, we express the estimates as a function of given points $k_0$ and $k_1$ as follows:
\begin{align}
	\label{eq_estimates_simplified_a}
	\hat{\alpha} &= F_{\alpha}(k_0,k_1)\\
	\label{eq_estimates_simplified_g}
	\hat{\gamma} &= F_{\gamma}(k_0,k_1)\\
	\label{eq_estimates_simplified_b}
	\hat{\beta} &= F_{\beta}(k_0,k_1,\hat{\alpha},\hat{\gamma})\\
	\label{eq_estimates_simplified_d}
	\hat{\delta} &= F_{\delta}(k_0,k_1,\hat{\alpha}),
\end{align}
where $\hat{\beta}$ and $\hat{\delta}$ additionally needs the information of the estimates $\hat{\alpha}$ and $\hat{\gamma}$.
Sometimes, the estimates can possibly outrange the parameter spaces $\alpha \in (0,2],\,\beta \in [-1,1]$, and $\gamma > 0$, especially when the true parameters are close to the borders.
In such cases, the parameters are set to the closet border, except for $\alpha$ and $\gamma$, the estimates are set no lower than 0.01.
Applications with other parameterizations use slightly different forms of CF, but the stable parameters are estimated essentially by the same procedure as explained above.
For the zero-parameterization, which is another common parameterization form, the CF is replaced to its corresponding form shown in equations~\eqref{eq:first2} and~\eqref{eq:parameterization_relation} for equations~\eqref{eq_absCF1} and ~\eqref{eq_cumulantfunction1} (or~\eqref{eq_cumulantfunction3}).
For parameterization with a different definition of the scaling parameter written as $c\,(=\gamma^\alpha)$~\cite{Nikias1995, Bibalan2017, Liu2018}, Bibalan et al. (2017)~\cite{Bibalan2017} presents an alternative procedure for the estimation.
They first directly obtain the scaling parameter $c$ from taking the absolute value of the empirical CF, or the real part of the cumulant function as
\begin{align}
	\label{eq_Bibalan_c}
	\hat{c} = -\ln |\hat{\varphi}(1)| = -\Re \left\{ \ln \hat{\varphi}(1) \right\}.
\end{align}
Next $\alpha$ is estimated as shown in equation~\eqref{eq_alpha_estimate}.
Then, the scale parameter in our criterion, $\hat{\gamma}$, is obtained as,
\begin{align}
	\label{eq_Bibalan_g}
	\hat{\gamma} = \exp\left( \frac{\ln \hat{c}}{\hat{\alpha}} \right).
\end{align}
The remaining parameters $\beta$ and $\delta$ are then estimated straightforwardly as similar to the case of the one-parameterization form.
By replacing $\hat{\gamma}^{\hat{\alpha}}$ to $\hat{c}$ in equations~\eqref{eq_beta_estimate} and~\eqref{eq_delta_estimate} (or equations~\eqref{eq_beta_estimate2} and~\eqref{eq_delta_estimate2}), and using the points $k_0$ and $k_1$ give the estimates.
%==============================================================================
%%%%%%%%%%%%%%%%%%%%%%%%%%%%%%%%%%%%%%%%%%%%%%%%%%%%
% 4. The Choice of Two Points for the characteristic function-based method
%%%%%%%%%%%%%%%%%%%%%%%%%%%%%%%%%%%%%%%%%%%%%%%%%%%%
\section{Proposed Approach for the Characteristic Function-based Method}
In this section, we make an improvement of the CF-based method by discussing how the points related to the CF should be chosen.
We propose a technique that provides a flexible selection of the points.
We also clarify the difference of how the points are selected between our proposal and the procedures in other existing CF-based methods.
%%%%%%%%%%%%%%%%%%%%%%%%%%%%%%%%%%%%%%%%%%%%%%%%%%%%
% 4.1 Inference of the first point $k_1$
%%%%%%%%%%%%%%%%%%%%%%%%%%%%%%%%%%%%%%%%%%%%%%%%%%%%
\subsection{Argument for the inference of point $k_1$}
Two positive points of the CF, $k_0$ and $k_1 (k_0 \neq k_1)$, are ought to be selected to identify all four parameter estimates.
As mentioned before in this paper, the absolute value of the CF in equation~\eqref{eq_absCF1} is independent of the skew and location parameters for any $k$, and provides information of $\alpha$ and $\gamma$.
When $k=1/\gamma$ is satisfied, the absolute value of the CF takes a constant value
\begin{align}
	\label{eq:propose1}
	\left| \varphi \left(1/\gamma \right) \right|=\mathrm{e}^{-1}.
\end{align}
The advantage of setting $k=1/\gamma$ as one of the candidate points is to reduce any estimation bias influenced by certain parameter values since we expect to get a constant estimate which is independent of all four parameters.
When $\gamma\gg1$, however, empirically obtained values can cause significant estimation errors for the scale parameter in equation~\eqref{eq:propose1}~\cite{Krutto2018,Paulson1975}.
Therefore, we first consider a temporary estimate of the scaling parameter, $\tilde{\gamma}$, just in case the data exhibits scale far from the standardized form ($\gamma=1$).

Take the natural logarithm of equation~\eqref{eq:propose1}.
The temporary estimate can be obtained by approximately solving the equation that numerically satisfies
\begin{align}
	\label{eq:propose2}
	\ln \left| \hat{\varphi} \left(1/\tilde{\gamma} \right) \right| \simeq -1,
\end{align}
using a simple one-dimensional search function~\cite{Brent2013}, or any other optimization procedure.
Our rough estimate $\tilde{\gamma}$ is then used for standardizing, or pre-standardizing, the candidate points.
Specifically, point $k_1$ is set to $1/\tilde{\gamma}$, where $\ln |\varphi(k_1)|$ empirically takes $-1$.

As explained above, pre-standardization is preferred especially when we suspect that datasets have too large or small scales.
Whenever a new set of points is required for the parameter estimation process, we conduct pre-standardization.
Point $k_1$ is replaced to $1/\check{\gamma}$, where $\check{\gamma}$ is the latest scaling parameter estimate available at that time.
%%%%%%%%%%%%%%%%%%%%%%%%%%%%%%%%%%%%%%%%%%%%%%%%%%%%
% 4.2 Inference of point $k_1$
%%%%%%%%%%%%%%%%%%%%%%%%%%%%%%%%%%%%%%%%%%%%%%%%%%%%
\subsection{Argument for the inference of point $k_0$}
For the argument of selecting point $k_0 > 0$, which is perhaps the most important proposal in our study.
We focus on the absolute value of the CF.
Bibalan et al. (2017) proposed to calculate the distance between two absolute values of CFs with different index parameters $\alpha$, the Gaussian case ($\alpha=2$) and the Cauchy case ($\alpha=1$)~\cite{Bibalan2017}.
They set $k_0 >0$ to the point which corresponds to the maximum distance and the other point to $k_1 = 1$.
Although the absolute CF changes depending on the index parameter $\alpha$, their approach considers a fixed distance and essentially chooses an identical point for any value of $\alpha \in (1,2]$.
In addition, the distance they consider does not account for the case of $\alpha \in (0,1]$.

Our approach is an extension of Bibalan et al. (2017), and provides a more generalized technique of selecting the points.
We deal with the problem that the distance between two absolute values of CFs can vary depending on the parameters.
The basic idea is to find the point where the absolute CF, $|\varphi(k; \alpha)|$, presents the {\it maximum sensitivity} with respect to $\alpha$.
In other words, we discuss the point where the distance between the absolute CF of index parameter $\alpha$, $|\varphi(k; \alpha,\beta,\gamma,\delta)|$, and the absolute CF of $\alpha + \Delta \alpha$, $|\varphi(k; \alpha+\Delta \alpha,\beta,\gamma,\delta)|$, shows the largest distance.
Such a point is considered as $k_0$ in our study.

To make our discussion more simple, we consider the absolute CF as a function of variable $\eta$:
\begin{align*}
	|\varphi(k;\alpha, \beta, \gamma, \delta)| = \exp(-\eta^\alpha),
\end{align*}
where $\eta=\gamma k\; (k>0)$ is a newly introduced variable which depends on $\gamma$ and $k$.
The distance can be expressed as $\left| \exp(-\eta^{\alpha+\Delta \alpha}) - \exp(-\eta^{\alpha}) \right|$.
The candidate point for $\eta_0=\gamma k_0$, where the maximum distance is achieved, can be calculated by
\begin{align}
	\label{eq:propose3}
	\frac{d}{d\eta} \left| \exp(-\eta^{\alpha+\Delta \alpha}) - \exp(-\eta^{\alpha}) \right| = 0,\; \eta>0.
\end{align}
Solving this equation for $\eta>0$ yields two solutions, $\eta \in(0,1/\gamma)$ and $\eta \in (1/\gamma, \infty)$.
For both points, the absolute value of CF shows the largest ratio of change in a local sense.
The smaller point $\eta \in(0,1/\gamma)$ is employed, because the distance at the smaller point tends to have larger values than that at the larger point $\eta \in (1/\gamma, \infty)$, which enables us to estimate $\alpha$ and $\gamma$ in a more desirable and informative manner.
Another reason is that smaller $|k|$ is preferred rather than larger $|k|$.
As $|k|\rightarrow 0$, the asymptotic variance of the empirical cumulant function decreases~\cite{Krutto2018}.
With empirical CF obtained by i.i.d. distributed datasets, the relation
\begin{align}
	\label{eq:propose_small_k}
	E\left[ \left | \varphi_N (k) \right |^2 \right ] = |\varphi(k)|^2 + \frac{1}{N} \left(1- |\varphi(k)|^2 \right),
\end{align}
holds~\cite{Kakinaka2020}, which implies that as $k$ becomes larger, the empirical absolute CF $|\varphi_N(k)|$ is likely to be subject to sample errors.
Thus, the smaller $\eta =\gamma k$ should be considered in this study.

The above discussion implies that $k$ should be set close to zero (but not at zero because then the CF takes a constant value and no information of the parameters will be provided).
But at the same time, the employed smaller point is standapart from zero to some extent, so that the empirical CF will be more or less exposed by sample errors.
Therefore, the choice of points derived from equation~\eqref{eq:propose3} is unsatisfactory, and hence the distance $\left| \exp(-\eta^{\alpha+\Delta \alpha}) - \exp(-\eta^{\alpha}) \right|$ should be modified.
To reduce the effect of sample errors, we introduce a weight function $w(\eta)$ that decreases monotonically as $\eta$ becomes larger (note that the introduced variable $\eta = \gamma k$ has a linear relationship with $k$).

Using the weight function $w(\eta)$, we now introduce a weighted distance $\left| \exp(-\eta^{\alpha+\Delta \alpha}) - \exp(-\eta^{\alpha}) \right| w(\eta)$ for $\eta>0$.
For convenience, we employ $w(\eta) = \exp{(-\tau|\eta|) }$, where $\tau>0$, since the CF exhibits an exponential form.
This choice leads to the association of the weighted distance with a statistical measure used for goodness-of-fit tests, developed by Matsui and Takemura (2008)~\cite{Matsui2008}.
They propose the following test statistic based on empirical CFs,
\begin{align}
	\label{eq:propose_matsui}
	\nonumber
	D_{N,\kappa} &:= N \int_{-\infty}^\infty \left| \hat{\varphi}(t) - \exp{(-|t|^{\alpha})} \right|^2 h(t) dt,\\
	h(t) &= \exp{(-\kappa |t|)}, \;\; \kappa>0,
\end{align}
where $h(t)$ is a monotonically decreasing weight function.
$D_{N,\kappa}$ denotes the weighted $L^2$-distance between the empirical CF and the symmetric standardized stable CF $\varphi(t;\alpha,0,1,0)$.
This weighted $L^2$-distance can be associated with the weighted distance we are considering now.

Taking the absolute value of a CF yields again a standardized form of a CF with $\beta=0$ and $\delta=0$:
\begin{align*}
	\exp(-\eta^\alpha) = |\varphi(k;\alpha,\beta,\gamma,\delta)| = \varphi(\eta;\alpha,0,1,0).
\end{align*}
Thus, the absolute values of CF with index parameter $\alpha$ and $\alpha+\Delta \alpha$ are equivalent to the symmetric standardized stable CFs, $\varphi(\eta;\alpha,0,1,0)$ and $\varphi(\eta;\alpha+\Delta \alpha,0,1,0)$, respectively.
The weighted $L^2$-distance between these CFs essentially coincides $D_{N,\kappa}$, when the weight function satisfies
\begin{align*}
	w(\eta) = \sqrt{h(\eta)},
\end{align*}
for $\eta>0$.
In this case, the difference between the CFs can be evaluated more accurately with the background of a meaningful measurement.
Following Matsui and Takemura (2008), the asymptotic distribution of $D_{N,\kappa}$ is numerically evaluated and the critical values of the test statistics are approximately obtained~\cite{Matsui2008}.
Through computational simulation, they provide evidence that the test is most powerful when $\kappa = 5.0$ ($h(\eta)=\exp(-5|\eta|)$), especially for heavy tailed distributions.
Thus, our choice of the weight function is $w(\eta)=\exp(-2.5|\eta|)$, since $\tau = \kappa/2$.
Other weight functions such as $w(\eta)=\exp{(-|\eta|)}$ and $w(\eta)=\exp{(-\eta^2)}$~\cite{Paulson1975, Heathcote1977} can be employed, but lacks a conclusive evidence for the use of these alternatives.

With the weight function, the candidate points $\eta>0$ are calculated by solving the following equation:
\begin{align}
\nonumber
	&g(\alpha,\eta) \\
	\label{eq:propose_g}
	&= \frac{d}{d\eta} \left \{\left( \exp(-\eta^{\alpha+\Delta \alpha}) - \exp(-\eta^{\alpha}) \right)\cdot \exp(-\tau \eta) \right \} = 0,
\end{align}
where $\tau=2.5$.
Then we have
\begin{align}
	\label{eq:propose_g_fin}
	\nonumber
	g(\alpha,\eta) &= (\alpha \eta^{\alpha-1}+\tau)\exp(-\eta^\alpha - \tau \eta)\\
	&- \left( (\alpha+\Delta\alpha)\eta^{\alpha+\Delta\alpha-1}+\tau \right)\exp(-\eta^{\alpha+\Delta\alpha}-\tau\eta).
\end{align}
For convenience, $\Delta \alpha$ is set to $0.01$ for all cases in this study.
Equation $g(\alpha,\eta)=0$ indicates the relationship between the index parameter $\alpha$ and point $\eta$ that exhibits the maximum rate of a change, or the maximum sensitivity, of the absolute CF with respect to $\alpha$.

\begin{figure}
 \centering
  \includegraphics[width=\linewidth]{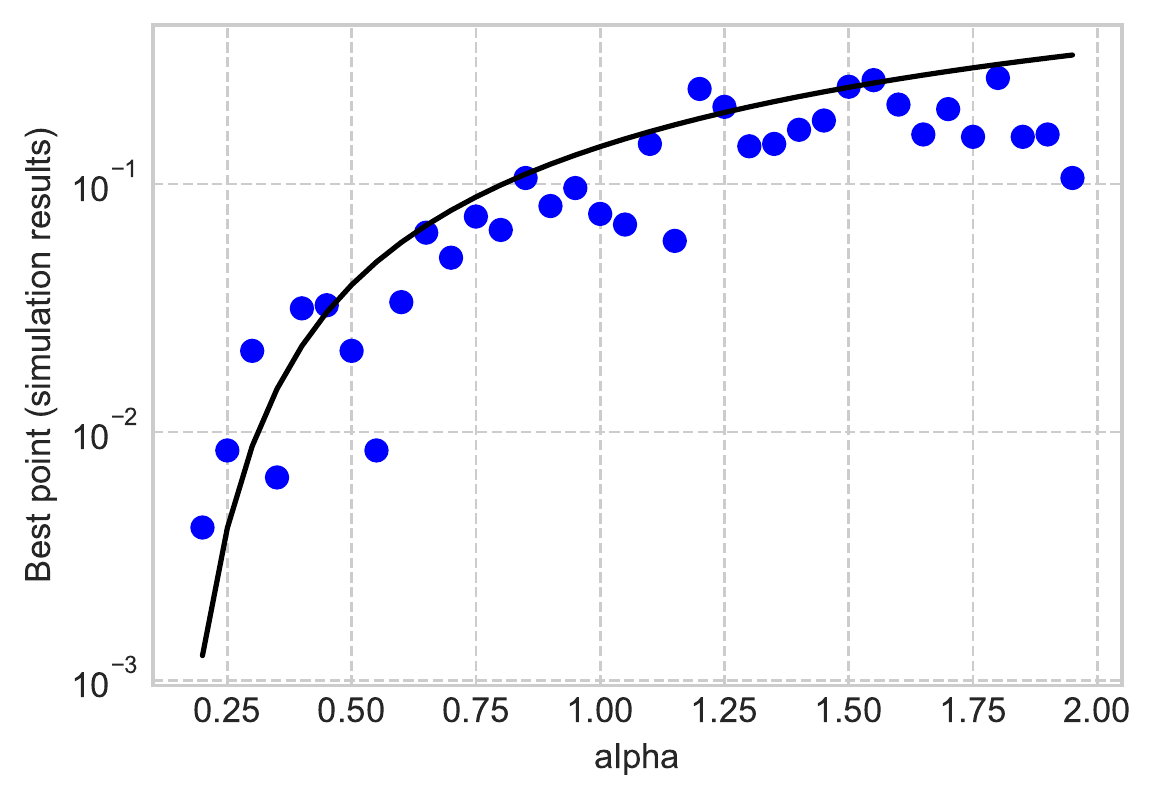}
  \caption{The theoretical relationship between $\alpha$ and $\eta$ based on our proposed selection approach, $g(\alpha, \eta)=0$ in equation~\eqref{eq:propose_g}, is shown in the solid black line.
  The blue plot shows the simulated results for the best point with the minimum MSE for $\alpha$ and $\beta$ over 100 simulations.
  We consider the MSE of $\alpha + \frac{1}{10}\beta$ because the accuracy of $\beta$ is generally worse roughly by ten times than the accuracy of $\alpha$, and also that $\beta$ estimates are usually susceptible to $\alpha$ estimates~\cite{McCulloch1986, Koutrouvelis1980, Bibalan2017}.
  The simulation is implemented for each value of $\alpha$ ranging within the parameter space of 0.2 to 1.95.}
  \label{fig:alpha_eta_ab}
\end{figure}

There could exist some relationship between $\alpha$ and $\eta$ since they are interrelated due to $g(\alpha,\eta)=0$.
When some estimate $\hat{\alpha}$ is given, the corresponding point is obtained by computing $\eta$ that satisfies $g(\hat{\alpha},\eta)=0$, and vice versa (the corresponding parameter $\alpha$ of a given point $\hat{\eta}$ can be calculated by computing the equation $g(\alpha,\hat{\eta})=0$).
As we have discussed previously in this subsection, we focus on the point closer (smaller) to zero out of the two candidates of the calculated points from equation~\eqref{eq:propose_g}.
Figure~\ref{fig:alpha_eta_ab} ascertains whether our approach of equation~\eqref{eq:propose_g} correctly estimates the parameters of stable distribution.
The model clearly characterizes the distinctive relationship between $\alpha$ and $\eta$, which are empirically verified via simulation using synthetic data generated from random stable variables~\cite{Weron1996}.
This indicates that our selection of points is valid for identifying desired points in the estimation process.

In practice, $\alpha$ is unknown.
Hence the selection of point $\eta_0=\gamma k_0$ is undecidable, so that the parameters for the stable law cannot be estimated directly.
To cope with this problem, we first aim to get a rough estimate of $\alpha$ calculated by using the temporary scale estimate $\tilde{\gamma}$.
The rough estimate is considered poor as the estimation method, but it plays a role in starting off the estimation process with reasonable initial values.
The accuracy of both points ($\eta_0=\gamma k_0$ and $\eta_1=\gamma k_1$) and the parameters ($\alpha,\beta,\gamma,\delta$) can be improved by alternating searches of $\alpha$ and $\eta$ from our relation model $g(\alpha, \gamma)=0$  several times to get sophisticated estimates.
With estimates $\eta_0$ and $\eta_1$, the four parameters are ultimately calculated.
%The next subsection presents our proposed algorithm for estimating stable laws, by making the most of the desire relationships between $\alpha$ and $\eta$.
%%%%%%%%%%%%%%%%%%%%%%%%%%%%%%%%%%%%%%%%%%%%%%%%%%%%
% 4.3 Estimation procedures
%%%%%%%%%%%%%%%%%%%%%%%%%%%%%%%%%%%%%%%%%%%%%%%%%%%%
\subsection{Estimation procedures}
Here we present our proposed algorithm for the estimation of all four parameters of stable laws by utilizing the relationship between $\alpha$ and $\eta$.
Regarding the fact that empirically obtained estimates occur substantial errors induced by $\gamma\gg1$, we conduct a pre-standardization with $k$ replaced to $\eta = \gamma k$.
Using the expressions of the estimates in equations~\eqref{eq_estimates_simplified_a}~\eqref{eq_estimates_simplified_g}~\eqref{eq_estimates_simplified_b}~\eqref{eq_estimates_simplified_d}, our algorithm is written as follows:
\begin{enumerate}[Step 1:]
	\item Compute a temporary estimate $\tilde{\gamma}_{\mathrm{temp}}$ from sample data $X_n\,(n=1,2,\ldots,N)$ that satisfies the equation,
	\begin{align*}
		\left. \ln \left |\frac{1}{N} \sum_{n=1}^{N} e^{i X_{n}/\gamma} \right| \,  \right |_{\gamma=\tilde{\gamma}_{\mathrm{temp}}} = -1.
	\end{align*}
	\item Set
	\begin{align*}
		\begin{cases}
		\tilde{k}_0 = \xi/\tilde{\gamma}_{\mathrm{temp}}\\
		\tilde{k}_1 = 1/\tilde{\gamma}_{\mathrm{temp}},
		\end{cases}
	\end{align*}
	where $\xi$ is any initial value of $\xi \in(0,1)$.
	\item Make a rough estimate of $\alpha$ and $\gamma$ from
	\begin{align*}
		\tilde{\alpha} &= F_\alpha (\tilde{k}_0, \tilde{k}_1)\\
		\tilde{\gamma} &= F_\gamma (\tilde{k}_0, \tilde{k}_1), 
	\end{align*}
	respectively, where $F_\alpha(\cdot,\cdot)$ and $F_\gamma(\cdot,\cdot)$ are given in equations~\eqref{eq_estimates_simplified_a} and~\eqref{eq_estimates_simplified_g}.
	\item Compute $\tilde{\eta}$ that satisfies $\left.g(\tilde{\alpha},\eta)\right|_{\eta=\tilde{\eta}}=0$, where $g(\cdot,\cdot)$ is given in equation~\eqref{eq:propose_g_fin}.
	\item Recalculate the points associated with $\tilde{\eta}$,
	\begin{align*}
		\begin{cases}
		\tilde{k}_0 = \tilde{\eta}/\tilde{\gamma}\\
		\tilde{k}_1 = 1/\tilde{\gamma},
		\end{cases}
	\end{align*}
	\item Estimate $\alpha$ and $\gamma$ as
	\begin{align*}
		\hat{\alpha} &= F_\alpha (\tilde{k}_0, \tilde{k}_1)\\
		\hat{\gamma} &= F_\gamma (\tilde{k}_0, \tilde{k}_1), 
	\end{align*}
	\item Compute $\tilde{\eta}$ that satisfies $\left.g(\hat{\alpha},\eta)\right|_{\eta=\hat{\eta}}=0.$
	\item Recalculate the points associated with $\hat{\eta}$,
	\begin{align*}
		\begin{cases}
		\hat{k}_0 = \hat{\eta}/\hat{\gamma}\\
		\hat{k}_1 = 1/\hat{\gamma},
		\end{cases}
	\end{align*}
	\item Finally, we estimate the parameters $\alpha$ and $\gamma$ as
	\begin{align*}
		\hat{\alpha} &= F_\alpha (\hat{k}_0, \hat{k}_1)\\
		\hat{\gamma} &= F_\gamma (\hat{k}_0, \hat{k}_1),
	\end{align*}
	\item Estimate the parameters $\beta$ and $\delta$ from the functions $F_\beta(\cdot,\cdot,\cdot,\cdot)$ and $F_\delta(\cdot,\cdot,\cdot)$ given in equations~\eqref{eq_estimates_simplified_b} and~\eqref{eq_estimates_simplified_d}, as
	\begin{align*}
		\hat{\beta} &= F_\beta (\hat{k}_0, \hat{k}_1, \hat{\alpha}, \hat{\gamma})\\
		\hat{\delta} &= F_\delta (\hat{k}_0, \hat{k}_1, \hat{\alpha}),
	\end{align*}
	which leads to the estimates of all four parameters of stable laws.
\end{enumerate}
%==============================================================================
%%%%%%%%%%%%%%%%%%%%%%%%%%%%%%%%%%%%%%%%%%%%%%%%%%%%
% 5. Simulation Assessments
%%%%%%%%%%%%%%%%%%%%%%%%%%%%%%%%%%%%%%%%%%%%%%%%%%%%
\section{Numerical Assessments}
In this section, we show numerical assessments for the estimation of stable laws.
We compare the performances of our proposal approach to other state-of-art approaches using the MSE and the KS-distance.
The comparison is studied for three approaches.
We focus on the approaches of characteristic function-based methods presented by Bibalan et al. (2017)~\cite{Bibalan2017} and Krutto (2018)~\cite{Krutto2018}.
We also compare with the traditional QM method~\cite{McCulloch1986, FamaRoll1971} explained in subsection 3.1, to provide a benchmark with a well-known criterion.
Note that all three approaches above exhibit closed-form expressions for all four estimates of stable parameters.

Bibalan et al. (2017) have shown that their approach generally outperforms other methods that yield a closed-form expression, such as the FLOM, the QM, and the MOLC~\cite{Bibalan2017}.
Krutto (2018) also compares the performances with several well-known methods and concludes that the method gives accurate estimates~\cite{Krutto2018}.
Since both of them belong to the family of the CF-based method, the selection of the points $k_0$ and $k_1$ plays an important role.
In Bibalan et al. (2017), $k_1$ is set to 1.
Point $k_0$ is always set to where the point shows the maximum distance between the absolute Gaussian CF and the absolute Cauchy CF, by using the estimates of $\gamma^\alpha$ which they are calculated beforehand.
It should be mentioned that the CF in this case poses a alternative definition of the scaling parameter, so we eventually obtain $\gamma$ in the last procedure in equation~\eqref{eq_Bibalan_g}.
On the other hand, Krutto (2018) suggests to employ two points that satisfies
\begin{align*}
    \begin{cases}
		\ln|\hat{\varphi}(k_0)| = -0.1\\
		\ln|\hat{\varphi}(k_1)| = -0.5,
	\end{cases}
\end{align*}
under empirical searches~\cite{Krutto2018}.
We examine the performance for each parameter of stable distribution in addition to the fit with the entire estimated stable distribution.
We also refer to the effects of sample sizes for each estimation method.
For all the simulations in this paper, we generate $L=500$ synthetic data of $N=10000$ i.i.d. random stable samples.
Synthetic random data sequences following a stable distribution can be generated by algorithms constructed by Chambers et al. (1976)~\cite{Chambers1976}, Weron (1996)~\cite{Weron1996}, and Umeno (1998)~\cite{Umeno1998}.
Umeno (1998) generates random stable variables based on the superposition of chaotic processes.
The classical method of Chambers et al. (1976) is widely known as the pioneer of all the methods, which the algorithm was reorganized and corrected later by Weron (1996).
Weron's algorithm is our choice of method, which is simple and is the fastest in calculation.
%%%%%%%%%%%%%%%%%%%%%%%%%%%%%%%%%%%%%%%%%%%%%%%%%%%%
% 5.1 Performance of parameter estimates
%%%%%%%%%%%%%%%%%%%%%%%%%%%%%%%%%%%%%%%%%%%%%%%%%%%%
\subsection{Performance of parameter estimates}
The performance of the estimated parameters are examined by the MSE criterion:
\begin{align*}
	\mathrm{MSE}(\theta) = \frac{1}{L} \sum_{l=1}^L \left( \theta-\hat{\theta}_l\right),
\end{align*}
where $\theta$ and $L=500$ is the parameter of stable laws and the number of times the simulation is implemented, respectively.
We calculate the MSE of all four parameters and evaluate each parameter individually.

Table~\ref{table_parameter} shows the simulation results of the MSE associated with the estimate bias for each parameter.
We consider the cases of parameters with $\alpha=0.5,1.5,1.8$ and $\beta=0, 0.5$, all with a standardized form of $\gamma=1$ and $\delta=0$.
Our proposed approach generally provides the most accurate estimation with the smallest MSE.
Especially for the index parameter $\alpha$ and $\delta$, our approach significantly improves the accuracy of the estimates.
Note that for the QM, the method has parameter restrictions of $\alpha\geq0.6$ and hence the cases with $\alpha$ smaller than 0.6 can not be implemented.
More detailed simulation results for different cases of parameters are shown in Figures~\ref{alpha_001},~\ref{beta_08},~\ref{beta_16} in Appendix A.
In particular, we show the cases of $S(\alpha,-0.1,1,0), S(0.8,\beta,1,0)$ and $S(1.6,\beta,1,0)$, with parameter values varying within the parameter ranges.
The results imply that for whatever parameter combination, our method generally outperforms the others with the highest accuracy.
Although we find that other methods sometimes show higher accuracy on either the parameter $\alpha$ or $\delta$ in cases of $0.6 \leq \alpha \leq 1.2$ in $S(\alpha, -0.1,1,0)$ in Figure~\ref{alpha_001} and $-0.3 \leq \beta \leq 0.3$ for $S(0.8,\beta,1,0)$ in Figure~\ref{beta_08}, our method appears to be powerful for estimating all four stable parameters.
\begin{table*}[t]
   \begin{center}
   \caption{
   Simulation results for the performance of all four stable law parameters.
   The comparison of the proposed method with other methods based on Bibalan et al., Krutto, and QM are examined for different values of $(\alpha, \beta)$ with a standardized form of $(\gamma, \delta)=(1,0)$.
   Absolute values of bias are given below the MSE in parentheses for all cases.
   The minimum value of MSEs among the methods are shown in bold for each case of parameters.
   }
   \begin{tabular*}{\linewidth}{@{\extracolsep{\fill}} l l l c c c c c c}\hline
   & & & \multicolumn{6}{c}{$\alpha$}\\ \cline{4-9}
   & & & \multicolumn{2}{c}{0.5} & \multicolumn{2}{c}{1.5} & \multicolumn{2}{c}{1.8}\\ \cline{4-9}
   & & & \multicolumn{2}{c}{$\beta$} & \multicolumn{2}{c}{$\beta$} & \multicolumn{2}{c}{$\beta$}\\ \cline{4-9}
   & & ($\times10^{-4}$) & 0 & 0.5 & 0 & 0.5 & 0 & 0.5\\ \hline\hline
   $\hat{\alpha}$ & proposed & MSE & \textbf{0.859} & \textbf{0.767} & \textbf{3.353} & \textbf{2.881} & \textbf{2.128} & \textbf{2.100}\\
   & & bias & (1.047) & (5.376) & (9.776) & (4.193) & (1.567) & (2.140)\\
   & Bibalan et al. & & 5.252 & 4.803 & 4.015 & 3.757 & 2.346 & 2.234\\
   & & & (8.435) & (4.880) & (2.793) & (16.51) & (2.994) & (1.710)\\
   & Krutto & & 1.387 & 1.429 & 4.958 & 4.604 & 2.816 & 2.728\\
   & & & (13.48) & (2.535) & (18.95) & (4.333) & (0.231) & (3.642)\\
   & QM & & --- & --- & 3.915 & 5.306 & 9.282 & 8.857\\
   & & & (---) & (---) & (4.732) & (16.75) & (16.63) & (16.97)\\\hline
   $\hat{\beta}$ & proposed & MSE & \textbf{6.867} & \textbf{7.522} & \textbf{11.54} & \textbf{11.55} & 40.68 & 48.78\\
   & & bias & (13.78) & (3.230) & (0.629) & (12.33) & (24.90) & (16.48)\\
   & Bibalan et al. & & 20.95 & 20.64 & 15.09 & 16.61 & 47.62 & 56.67\\
   & & & (19.32) & (5.274) & (17.51) & (4.882) & (36.72) & (19.40)\\
   & Krutto & & 11.66 & 12.64 & 15.71 & 15.18 & \textbf{37.05} & \textbf{42.97}\\
   & & & (0.736) & (3.166) & (9.488) & (3.711) & (7.387) & (29.83)\\
   & QM & & --- & --- & 11.59 & 13.01 & 61.39 & 162.3\\
   & & & (---) & (---) & (6.575) & (64.02) & (3.764) & (373.2)\\\hline
   $\hat{\gamma}$ & proposed & MSE & 15.95 & \textbf{13.20} & \textbf{1.444} & 1.396 & \textbf{0.842} & 0.857\\
   & & bias & (14.74) & (20.28) & (5.552) & (3.004) & (0.748) & (9.113)\\
   & Bibalan et al. & & \textbf{13.66} & 13.29 & 1.450 & \textbf{1.386} & 0.845 & \textbf{0.854}\\
   & & & (24.70) & (44.02) & (5.306) & (3.741) & (0.984) & (9.016)\\
   & Krutto & & 31.33 & 32.42 &	1.910 &	1.841 & 0.895 & 0.938\\
   & & & (48.27) & (25.64) & (13.73) & (4.923) & (1.007) & (9.917)\\
   & QM & & --- & --- & 1.613 & 1.989 & 1.483 & 1.518\\
   & & & (---) & (---) & (11.55) & (27.58) & (9.162) & (20.74)\\\hline
   $\hat{\delta}$ & proposed & MSE & \textbf{10.80} & \textbf{14.25} & \textbf{8.401} & \textbf{10.27} & \textbf{3.147} & 3.428\\
   & & bias & (11.90) & (33.02) & (10.70) & (1.020) & (0.243) & (2.800)\\
   & Bibalan et al. & & 30.86 & 35.41 & 10.72 & 13.43 & 3.497 & 3.965\\
   & & & (15.92) & (20.27) & (26.94) & (8.638) & (3.954) & (3.118)\\
   & Krutto & & 61.87 & 88.23 & 9.796 & 12.00 & 3.151 & \textbf{3.275}\\
   & & & (23.49) & (56.67) & (4.332) & (10.58) & (2.116) & (4.712)\\
   & QM & & --- & --- & 9.394 & 11.68 & 3.710 & 3.815\\
   & & & (---) & (---) & (14.87) & (46.61) & (1.920) & (32.97)\\
   \hline
   \end{tabular*}
   \label{table_parameter}
   \end{center}
\end{table*}
\subsection{Performance of the estimated distribution}
Next, we examine the performance of estimating stable laws from a different perspective; evaluation of the entire distribution.
We use the KS distance expressed as
\begin{align*}
	D = \max_{x}|P(x)-\hat{P}(x)|,
\end{align*}
which represents the maximum distance between two distributions in terms of the CDF.
Here $P(x)$ and $\hat{P}(x)$ denotes the empirically obtained CDF, and the theoretical estimated CDF, respectively.
The standard density and distribution functions of stable distributions are numerically derived approximately by implementing the Fourier integral formulas~\cite{Zolotarev1986, Nolan1997}, which are available in package {\it libstable} that provides good approximation values~\cite{Roy2017}.
KS distance is one of the most major standards for numerical assessments when discussing stable laws.
We set aside any issues related to numerical approximations of stable distributions, so that we can focus on the performance between the methods.
The root mean square (RMS) of the KS distance is used for the numerical assessment to make the small differences of the comparison results more apparent.
\begin{figure*}[pt]
\centering
  \subfigure[RMS of KS distances for $S(\alpha,0.1,1,0)$]{\includegraphics[clip, width=0.35\linewidth]{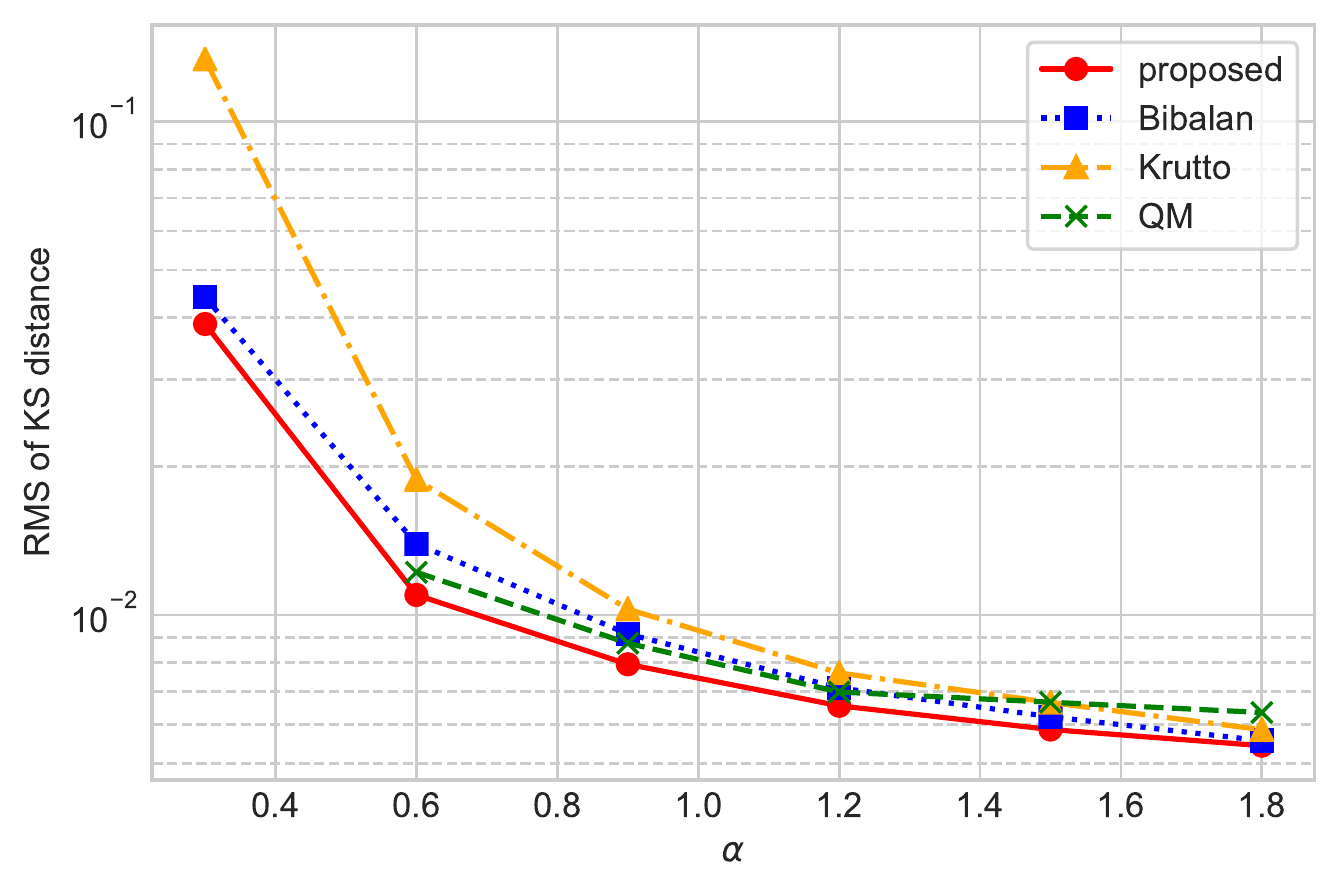}}
  \subfigure[RMS of KS distances for $S(1.7,\beta,1,0)$]{\includegraphics[clip, width=0.35\linewidth]{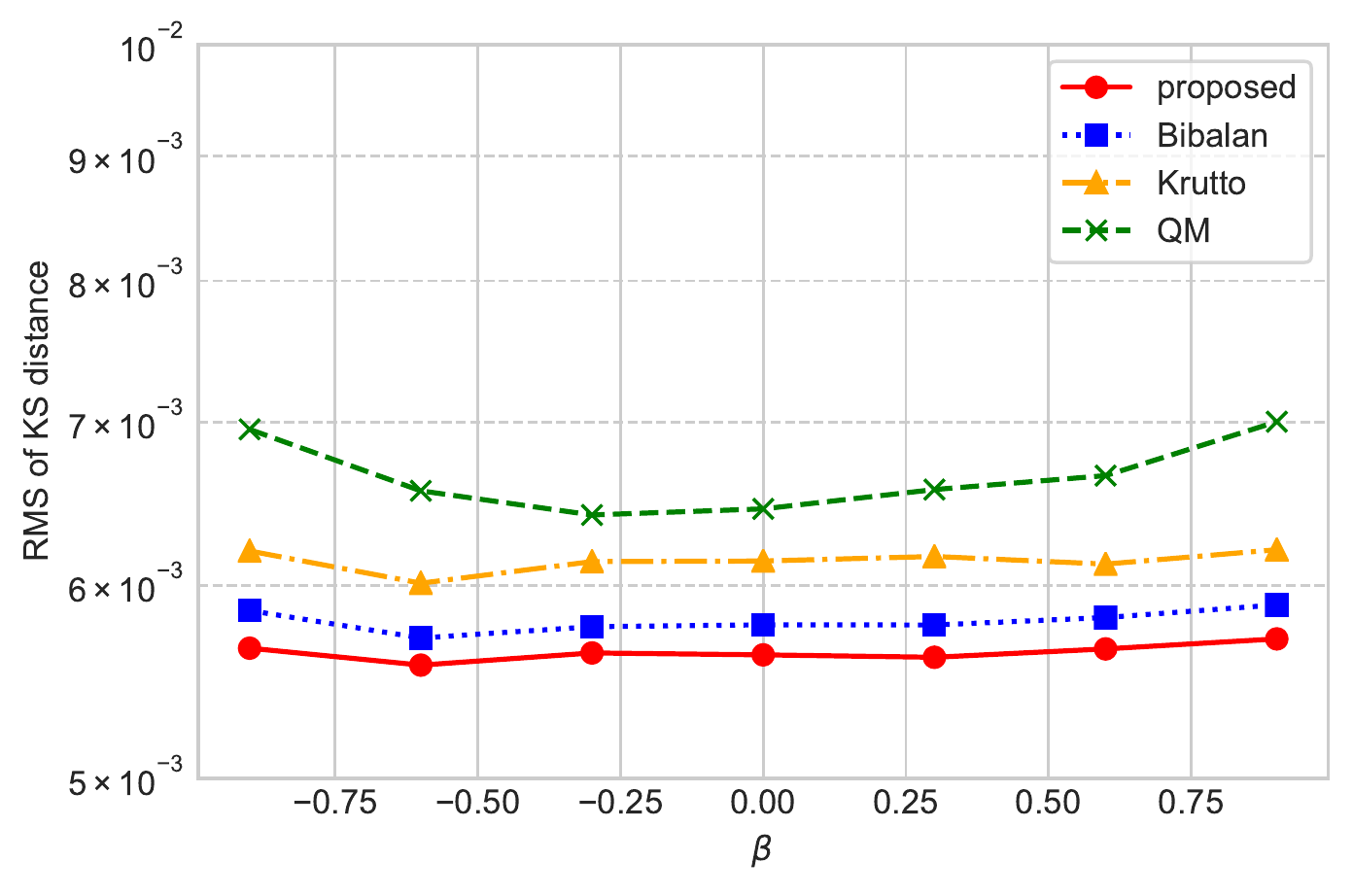}}\\
  \subfigure[RMS of KS distances for $S(1.3,0.2,\gamma,0)$]{\includegraphics[clip, width=0.35\linewidth]{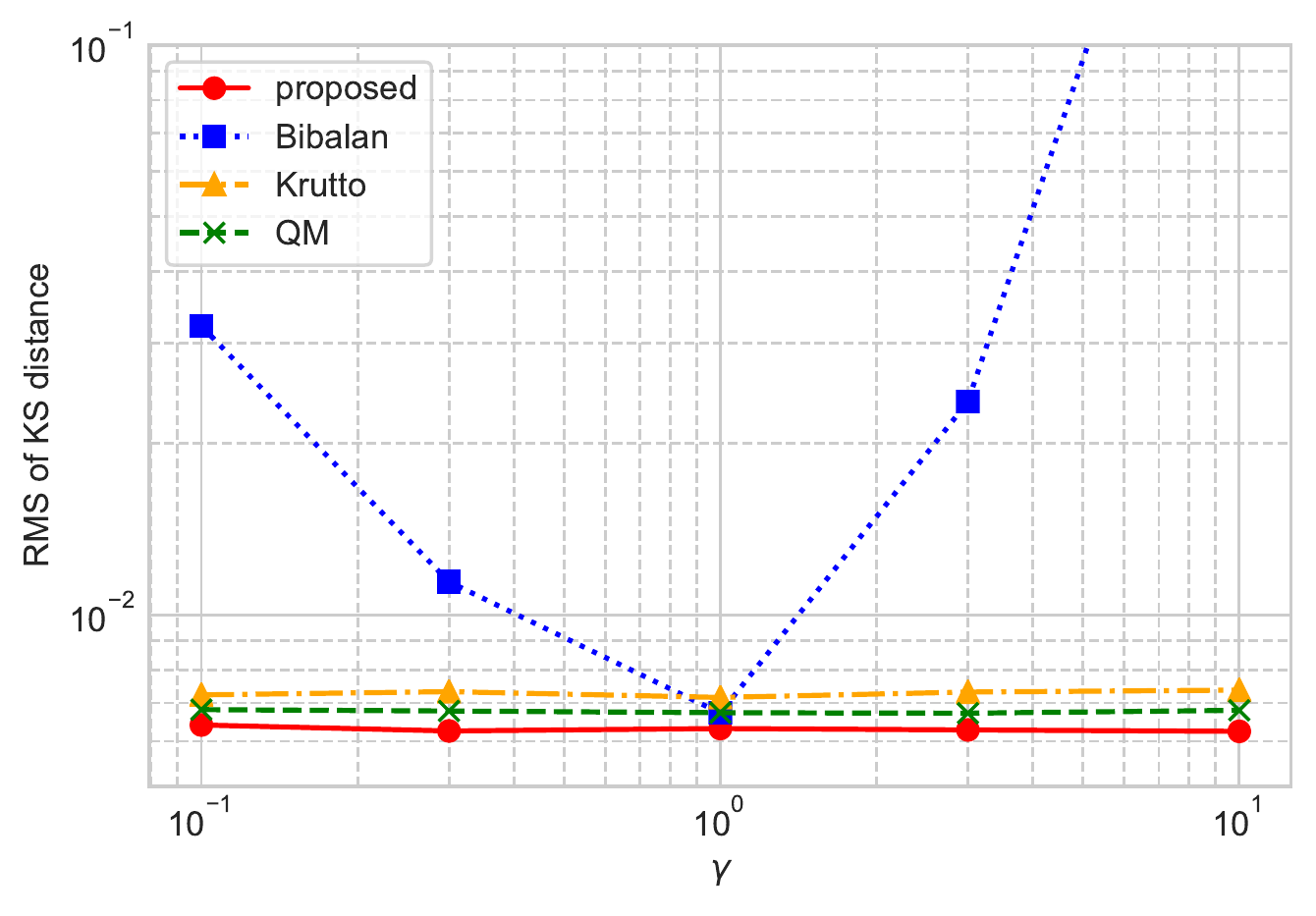}}
  \subfigure[RMS of KS distances for $S(0.7,-0.4,1,\delta)$]{\includegraphics[clip, width=0.35\linewidth]{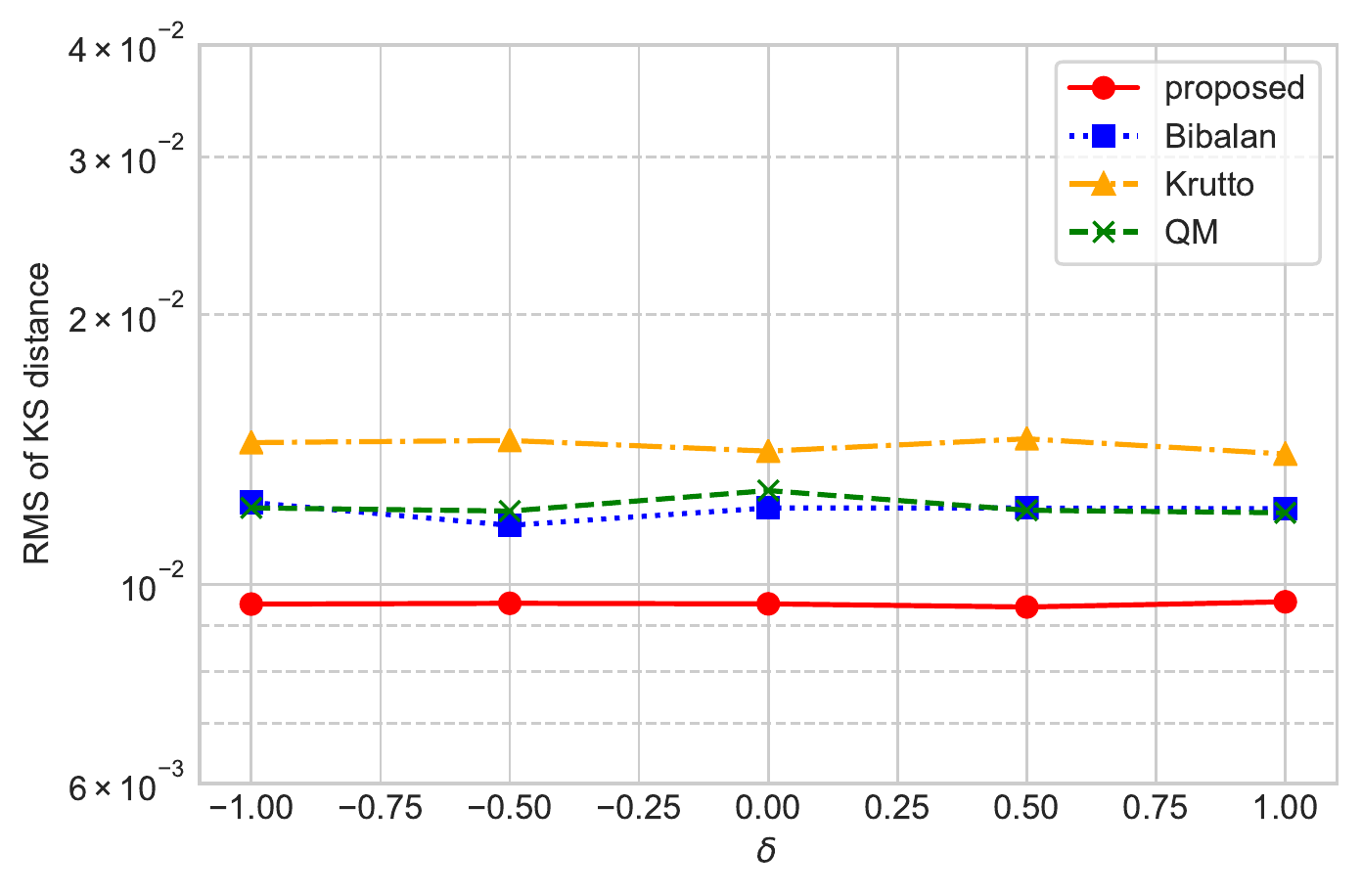}}
  \caption{
  Comparison of the KS distances for the methods based on the proposed approach, Bibalan et al.'s approach, Krutto's approach, and the QM method.
  The RMS values of KS distances are studied for several cases of stable distributions with parameters ranging within its parameter range ($N=10000, L=500$).
   }
   \label{ks_distance_method_comparison}
\end{figure*}
\begin{figure*}[pt]
\centering
  \subfigure[$\mathrm{MSE}(\alpha)$ for cases of $S(1.4,0.2,1,0)$]{\includegraphics[clip, width=0.35\linewidth]{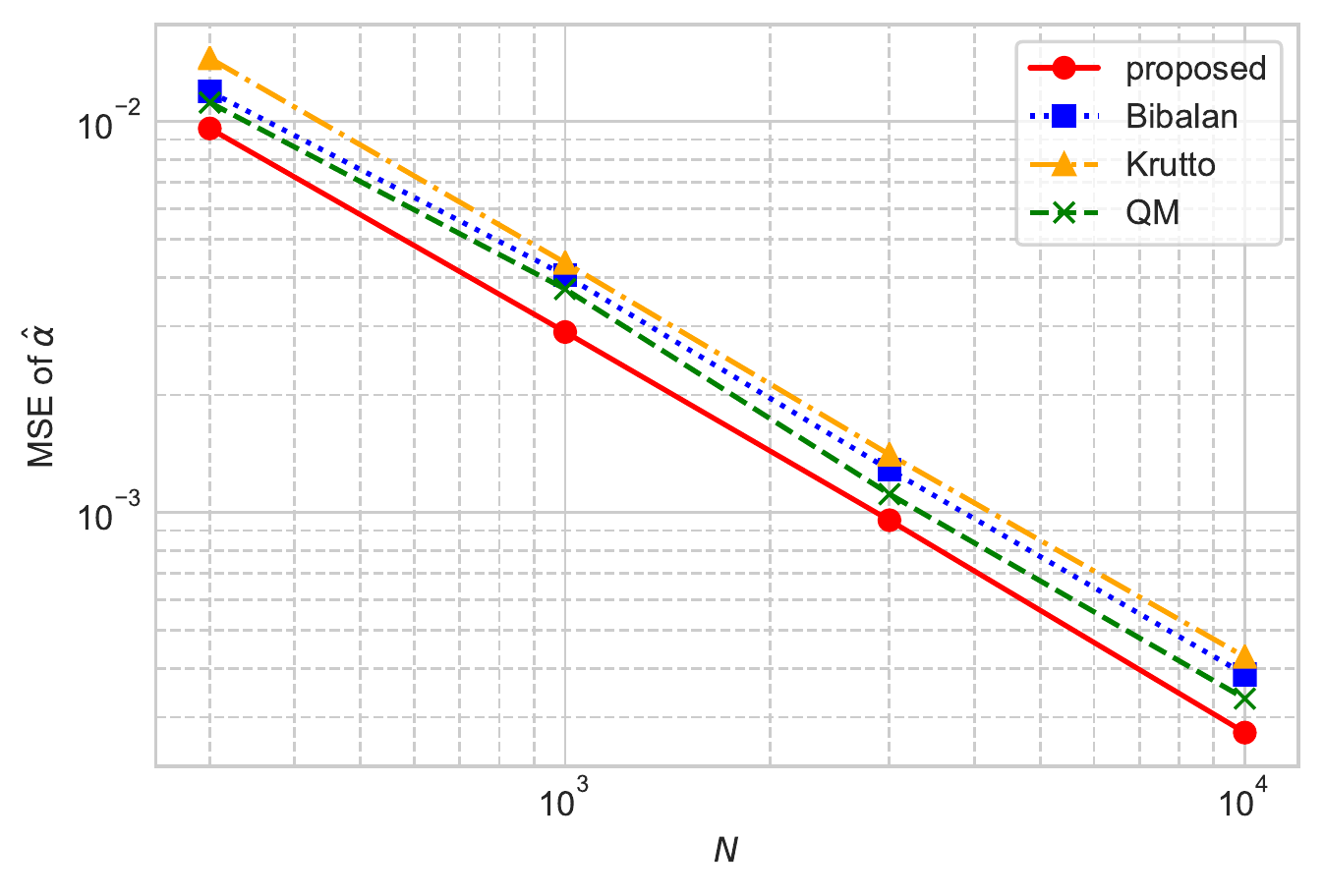}}
  \subfigure[$\mathrm{MSE}(\beta)$ for cases of $S(1.4,0.2,1,0)$]{\includegraphics[clip, width=0.35\linewidth]{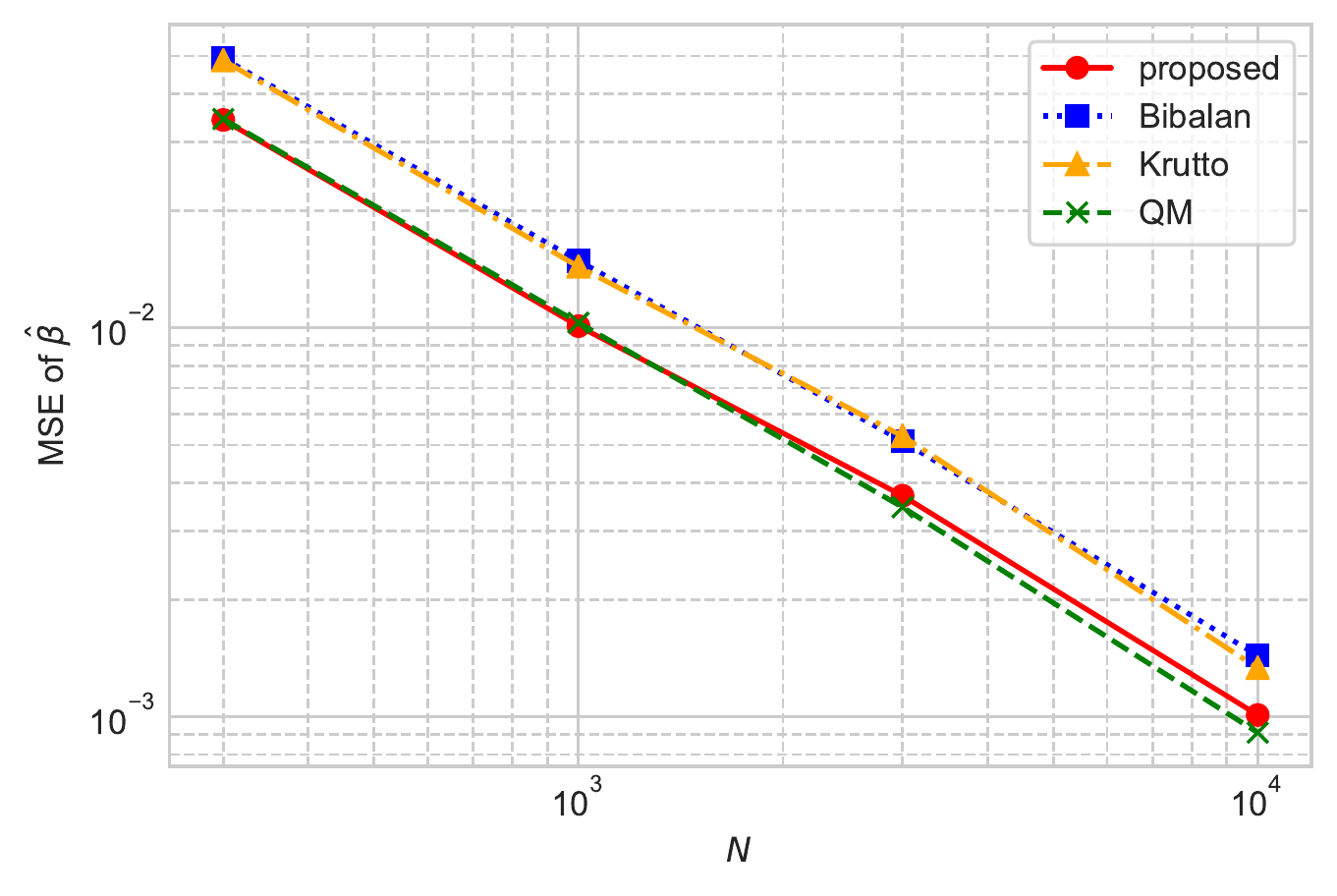}}
  \subfigure[$\mathrm{MSE}(\gamma)$ for cases of $S(1.4,0.2,1,0)$]{\includegraphics[clip, width=0.35\linewidth]{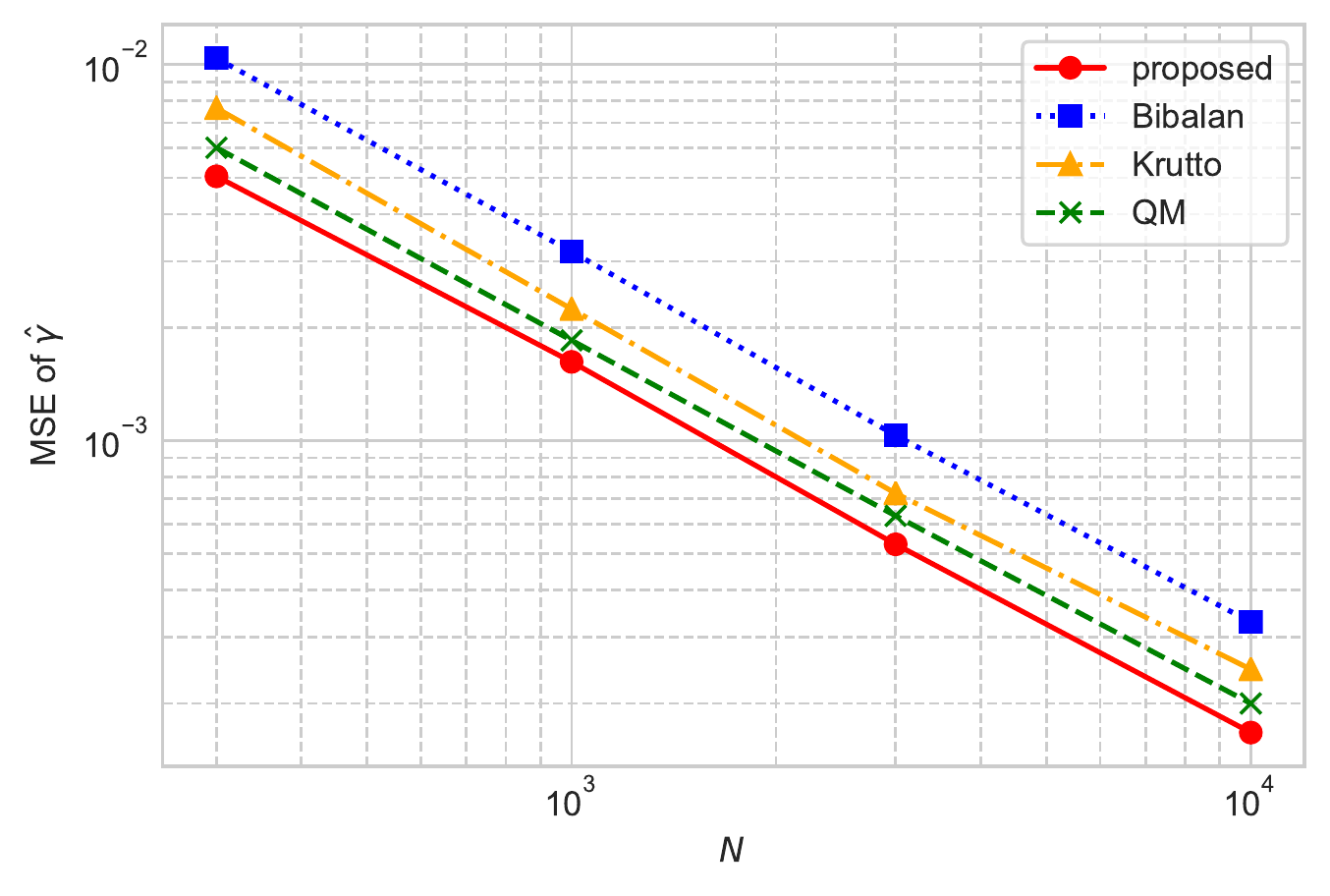}}
  \subfigure[$\mathrm{MSE}(\delta)$ for cases of $S(1.4,0.2,1,0)$]{\includegraphics[clip, width=0.35\linewidth]{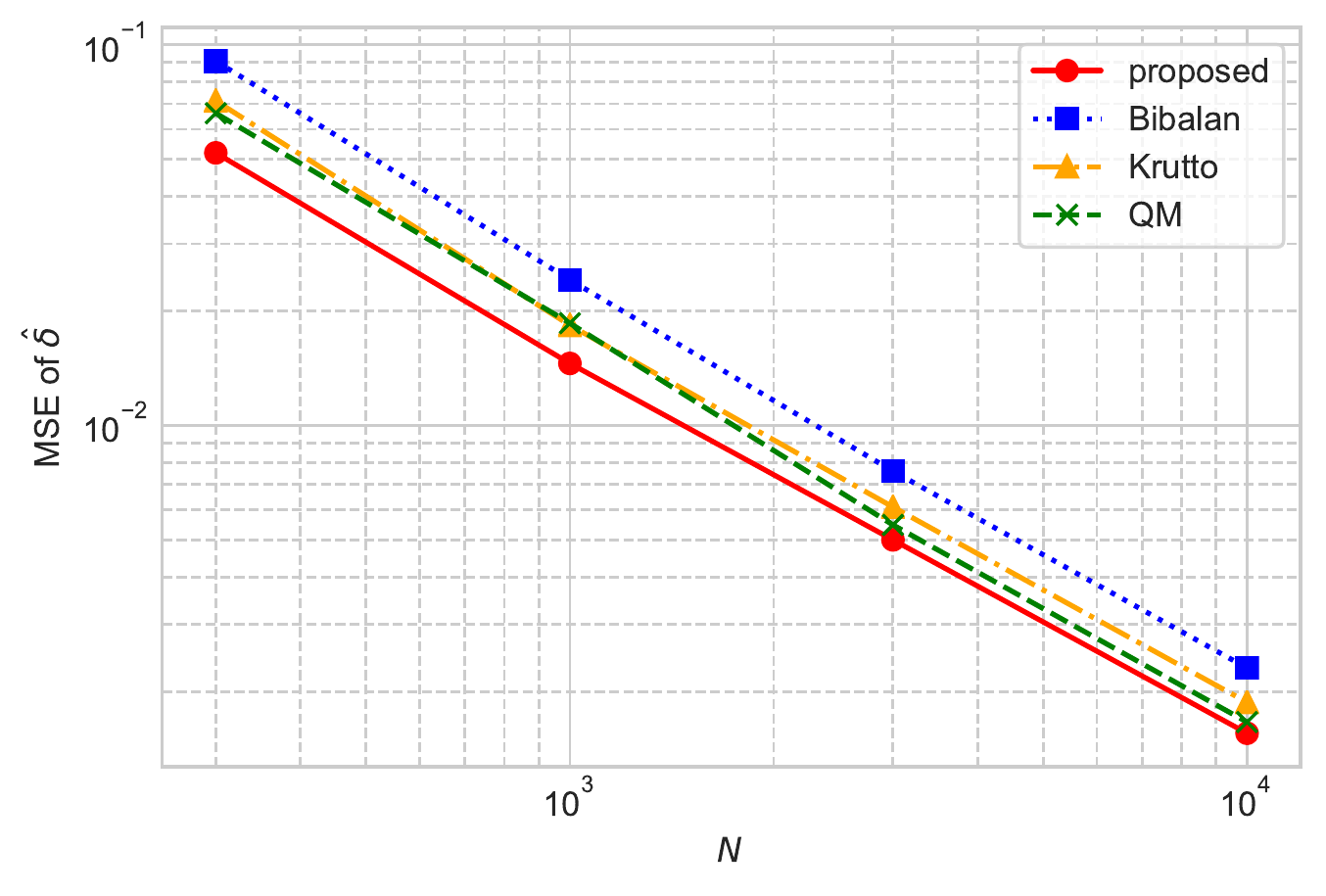}}
  \caption{
  Comparison of the MSE for the methods based on the proposed approach, Bibalan et al.'s approach, Krutto's approach, and the QM method with different values of sample sizes $N=300, 1000, 3000, 10000$.
  The MSE values of each stable parameter are studied for cases of $S(1.4,0.2,1,0)$ over $L=500$ synthetic datasets.
   }
   \label{MSE_N}
\end{figure*}
\begin{table*}[t]
\begin{center}
 \caption{Basic statistics of USDJPY and WTI return time series with time intervals of 1-hour and one day, respectively. Mean is the average of the return time series, SD is the standard deviation, and N is the number of sample sizes.
 }
 \begin{tabular*}{0.8\linewidth}{@{\extracolsep{\fill}} l c c c c c c c}\hline
 & Mean & SD & Skew & Kurt & Min & Max & N \\ \hline \hline
 USDJPY & 1.027$\times10^{-5}$ & 0.0062 & -0.0531 & 4.7880 & -0.0384 & 0.0550 & 4190\\
 WTI & -7.312$\times10^{-6}$ & 0.0041 & 0.5900 & 23.945 & -0.0576 & 0.1068 & 54356\\
 \hline
 \end{tabular*}
 \label{tb:basic_statistics}
 \end{center}
\end{table*}
Figure~\ref{ks_distance_method_comparison} shows the simulation results of the KS distance for several cases of stable distributions; $S(\alpha,0.1,1,0)$, $S(1.7,\beta,1,0)$, $S(1.3,0.2,\gamma,0)$, and $S(0.7,-0.4,1,\delta)$.
The RMS of the KS distance is calculated for each case with various values of parameters ranging within parameter ranges.
We find in Figure~\ref{ks_distance_method_comparison} (c) that the estimation for the scaling parameter $\gamma \neq 1$ poses significant estimation errors.
This is caused by the effect of sample errors induced by the scaling parameter $\gamma$ far from the standardized form, as shown in equation~\eqref{eq:propose1}.
On the other hand, our proposed method achieves the smallest value of KS distances for all cases of parameter combinations.
This proves that we are also successful in improving the estimation of the entire stable distribution.
%%%%%%%%%%%%%%%%%%%%%%%%%%%%%%%%%%%%%%%%%%%%%%%%%%%%
% 5.3 Effect of sample size
%%%%%%%%%%%%%%%%%%%%%%%%%%%%%%%%%%%%%%%%%%%%%%%%%%%%
\subsection{Effect of sample size}
Needless to say, the accuracy of the estimation method strongly depends on the number of samples.
Larger sample sizes give more information of the dataset whereas smaller sample sizes have only little information making it challenging to detect the true values.
We examine the effect of sample size by comparing the performance among the estimation methods.
Figure~\ref{MSE_N} displays the MSE of each parameter of stable distribution as the sample size $N$ changes from 300 to 10000.
The study is examined for the case of $S(1.4,0.2,1,0)$.
The MSE simulated by means of our method decreases with the order $\mathcal{O}(1/N)$ while the MSE simulated by means of other representative methods also exhibited similar behaviors of order.
Our proposed approach offers the best performance except for the location parameter $\delta$, where the QM method sometimes give more accurate estimates for large datasets.
%%%%%%%%%%%%%%%%%%%%%%%%%%%%%%%%%%%%%%%%%%%%%%%%%%%%
% 6. Application to Financial Empirical Data
%%%%%%%%%%%%%%%%%%%%%%%%%%%%%%%%%%%%%%%%%%%%%%%%%%%%
\section{Application to Financial Empirical Data}
This section shows application of the proposed estimation method to real financial data.
We provide several empirical studies to present that our proposed approach is appliable for a wide range of empirical analysis in finance.
\begin{table}[t]
\begin{center}
 \caption{Parameters of the fitted stable distribution for daily return time series of USDJPY exchange rate (2004/01/05-2019/12/31) and KS-distance calculated based on several estimation methods ($N=4190$).
 }
 \begin{tabular*}{\linewidth}{@{\extracolsep{\fill}} l c c c c c}\hline
 method & $\alpha$ & $\beta$ & $\gamma$ & $\delta$ & KS\\ \hline \hline
 proposed & 1.708 & -0.121 & 0.0035 & -0.00004 & \textbf{0.0214}\\
 Bibalan et al. & 1.884 & -0.261 & 0.0039 & -0.00002 & 0.0396\\
 Krutto & 1.767 & -0.138 & 0.0036 & -0.00004 & 0.0279\\
 QM & 1.584 & -0.064 & 0.0034 & -0.00012 & 0.0216\\
 \hline
 \end{tabular*}
 \label{tb:dowjones}
 \end{center}
\end{table}
\begin{table}[t]
\begin{center}
 \caption{Parameters of the fitted stable distribution for 1-hour return time series of WTI crude oil futures market (2010/11/14-2019/12/31) and KS-distance calculated based on several estimation methods ($N=54356$).
 }
 \begin{tabular*}{\linewidth}{@{\extracolsep{\fill}} l c c c c c}\hline
 method & $\alpha$ & $\beta$ & $\gamma$ & $\delta$ & KS\\ \hline \hline
 proposed & 1.357 & -0.045 & 0.0015 & -0.00007 & \textbf{0.018}\\
 Bibalan et al. & 1.846 & -0.012 & 0.0024 & -0.00002 & 0.088\\
 Krutto & 1.487 & -0.071 & 0.0017 & -0.00007 & 0.036\\
 QM & 1.260 & -0.031 & 0.0015 & -0.00009 & 0.019\\
 \hline
 \end{tabular*}
 \label{tb:wti}
 \end{center}
\end{table}

Asset price returns in various financial markets tend to show interesting properties of stable laws ever since Mandelbrot (1963) first revealed that stable distribution fits cotton price returns better than the classical Gaussian distribution~\cite{Mandelbrot1963}.
This argument have attracted attention to identifying price behaviors in many financial fields such as equities~\cite{Fama1965, MantegnaStanley1995, Xu2011}, price consumer index inflation~\cite{Chronis2016}, metal markets~\cite{Krezolek2012}, oil markets~\cite{Yuan2014}, and Cryptocurrency markets~\cite{Kakinaka2020}.
We investigate return distributions of the Japanese Yen currency exchange rate in terms of the US dollar (USDJPY) and the West Texas Intermediate (WTI) crude oil futures market, both of which are potent indices in finance.
The basic statistics of the indices are provided in Table~\ref{tb:basic_statistics}.
We explore both cases of common daily analysis and high-frequency data analysis.
In particular, we use daily and one-hour return time series for the USDJPY and the WTI market, respectively.
Since the scale of returns for both cases are too small for the method based on Bibalan et al. (2017) to give plausible estimates, we do a pre-standardization process beforehand.
We multiply returns by 100 and after the estimation the parameters $\gamma$ and $\delta$ are adjusted by dividing them by 100.
Table~\ref{tb:dowjones} presents the estimates of the fitted stable distribution associated with the KS-distance between the empirical distribution and the estimated stable distribution for USDJPY, calculated based on four controversial estimation methods.
Our primary focus is on the KS-distance value.
The results show that the estimated distribution based on our proposed method presents the smallest value among other estimation methods.
The smallest KS-distance implies that our method exhibits stable laws that best describes the observed data.
Parameter estimates and the distance measure for the WTI market are shown in Table~\ref{tb:wti}.
The result indicates that the outstanding performance of our method also holds for high-frequency data with the lowest KS-distance.
What makes the development of the estimation method a crucial matter is that the parameter estimates can differ so much among the methods when applied to empirically observed data, even for large datasets.
We find in Table~\ref{tb:wti} that the estimate of $\alpha$ marks a low 1.260 based on the QM method whereas Bibalan et al.'s method presents 1.846, which the value differs quite a lot between the methods in spite of the large sample size of dataset with $N=54356$.
A method that accomplishes the inference of the closest distribution or set of parameters provides a more reliable model.
Hence, our proposed estimation approach play a significant role as a tool for modeling with stable laws.
%%%%%%%%%%%%%%%%%%%%%%%%%%%%%%%%%%%%%%%%%%%%%%%%%%%%
% 7. Conclusion
%%%%%%%%%%%%%%%%%%%%%%%%%%%%%%%%%%%%%%%%%%%%%%%%%%%%
\section{Conclusion}
This paper has proposed a new approach for estimating stable laws and applied this approach to the exploration of price behaviors in financial markets.
Our new technique is developed under the method of moments, which is one of the widely known CF-based methods that require the choice of appropriate momental points.
The points necessary for the estimation process are flexibly chosen, as the estimation accuracy of stable laws depends heavily on their true parameter values.
We have focused on the fact that the index parameter $\alpha$ and the desired momental points exhibit a distinctive relationship, which is a new perspective in the literature.
This relation is modelled as $g(\alpha, \eta)=0$, based on the idea of employing points $\eta$ at which the weighted absolute values of the CF present the maximum sensitivity.
To detect appropriate points, we have suggested a procedure relying on the combination of empirical searches and algorithmic approaches.
The advantage of employing these points is that the parameters of stable laws can be estimated in a more precise manner while remaining straightforwardly the implementation of the method.
The relative performance of the parameter estimates is benchmarked against other existing methods, specifically the QM and the methods of Bibalan et al. (2017) and Krutto (2018), through simulation studies in terms of the MSE and KS-distance criteria.
The results have implied that our method is the {\it most powerful} with the best performance.
Our approach assures that the estimates of all four parameters of stable laws present a closed-form expression without any restrictions on parameter ranges, making the method significantly practical.
We have also explored the behaviors of price fluctuations in several financial markets to show that our method is applicable for empirical financial studies.
For the USD-JPY exchange rate and the WTI crude oil future price, our method supports stable laws with the highest performance among all the other methods discussed in this paper.
This would motivate us to further develop analytical methods for examining stable laws, as well as to further investigate various features of financial markets.

%\begin{acknowledgment}
%\acknowledgment
%\end{acknowledgment}
\appendix
%Figures
\section{Figures of simulation results}%%%%%%%%%%%%%%%%%%%%
We show in this section some of the additional simulation results examined for checking the performance of the parameter estimates.
Each of the four parameters of stable laws are studied for various cases of parameter combinations.
The results imply that for most cases, our proposed approach based method leads to improve the accuracy of the estimates.
We find that the state of performance is also consistent with all four parameters, outperforming the other existing methods.
%
%\begin{comment}
\begin{figure*}[h]
 \centering
  \subfigure[$\mathrm{MSE}(\alpha)$ for cases of $S(\alpha,-0.1,1,0)$]{\includegraphics[width=0.35\linewidth]{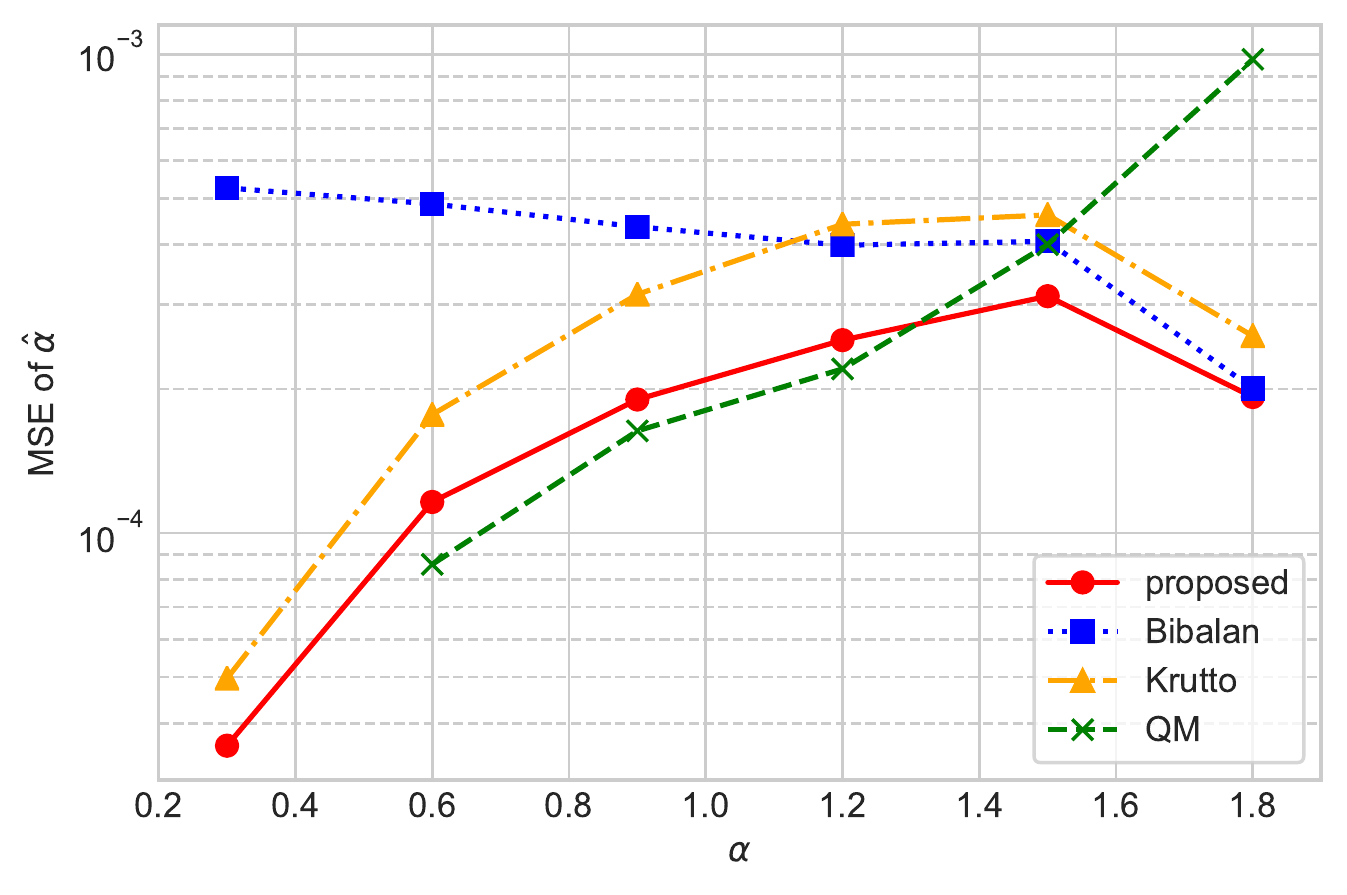}}
  \subfigure[$\mathrm{MSE}(\beta)$ for cases of $S(\alpha,-0.1,1,0)$]{\includegraphics[width=0.35\linewidth]{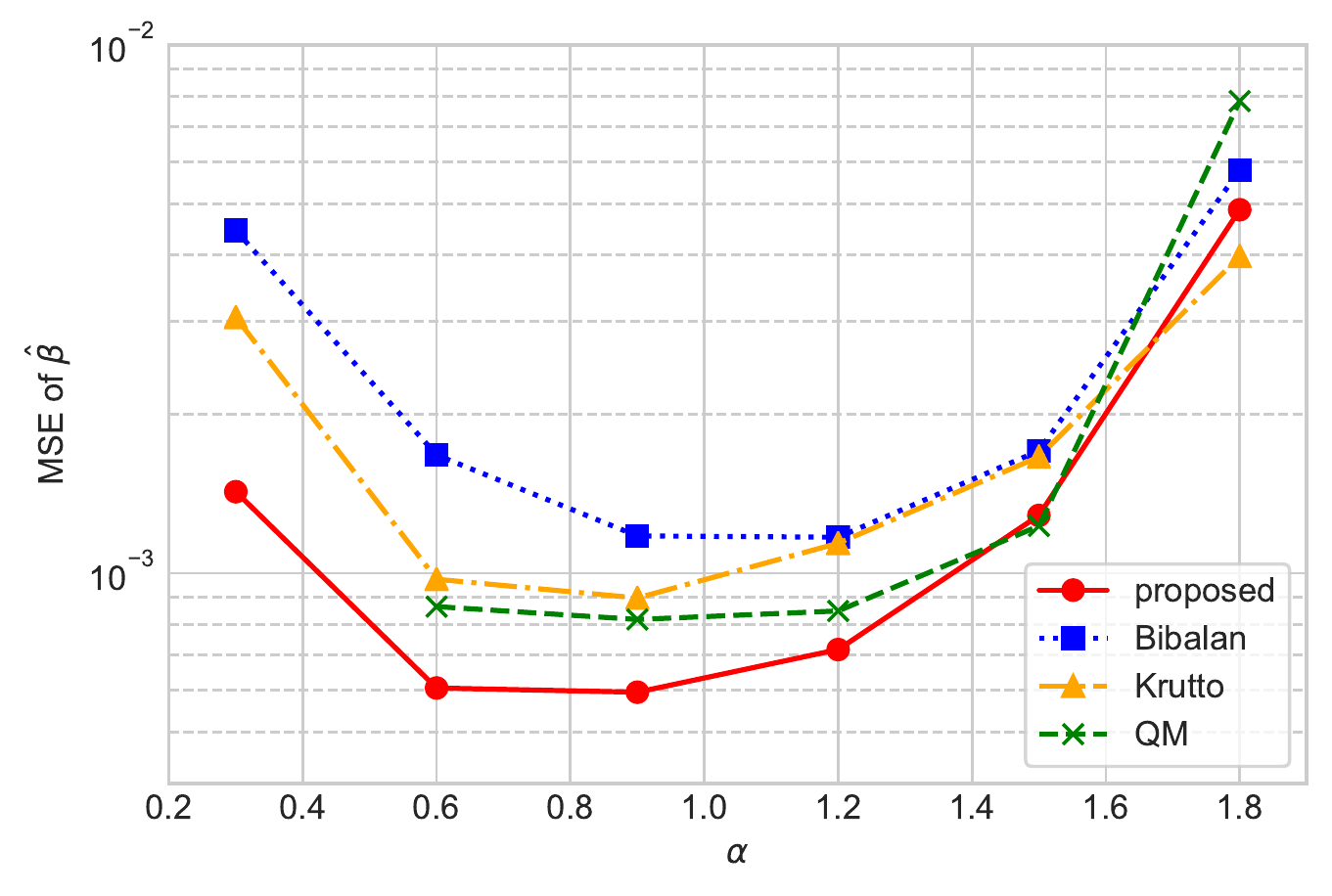}}
  \subfigure[$\mathrm{MSE}(\gamma)$ for cases of $S(\alpha,-0.1,1,0)$]{\includegraphics[width=0.35\linewidth]{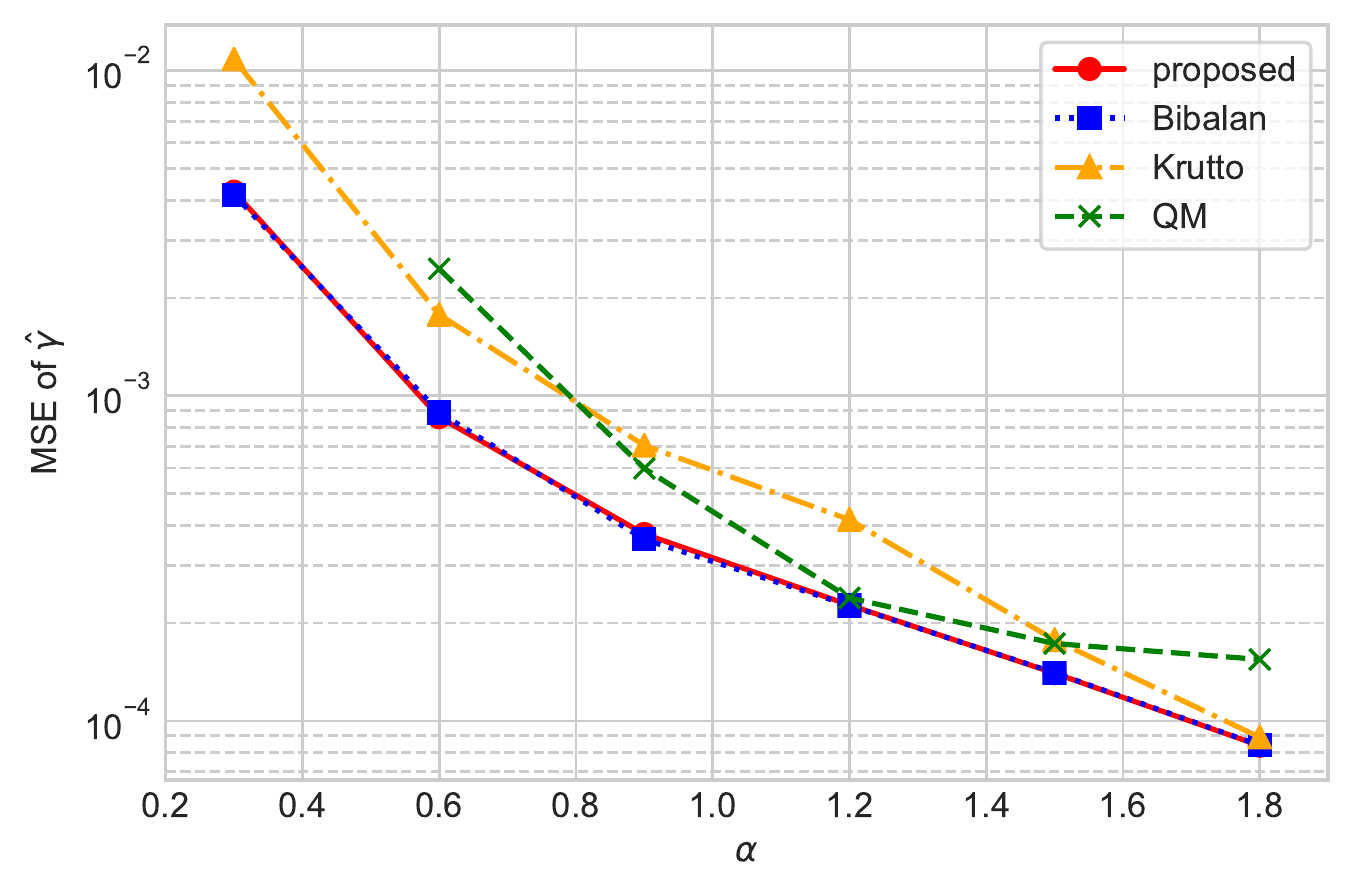}}
  \subfigure[$\mathrm{MSE}(\delta)$ for cases of $S(\alpha,-0.1,1,0)$]{\includegraphics[width=0.35\linewidth]{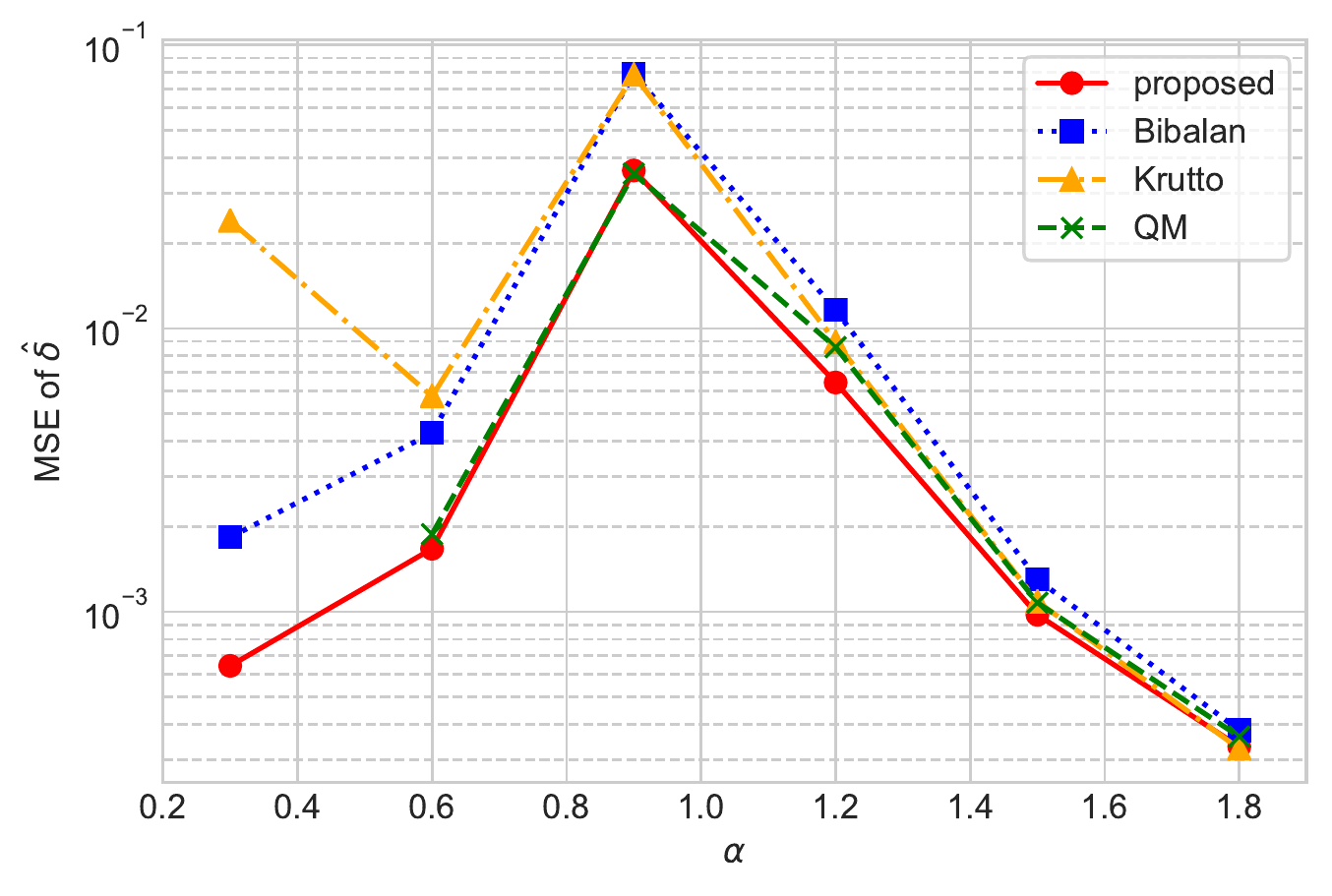}}
  \caption{
  Comparison of the MSEs for the methods based on the proposed approach, Bibalan et al.'s approach, Krutto's approach, and the QM method.
  The MSEs of each stable parameter are studied for cases of parameters $\beta=-0.1, \gamma=1$, and $\delta=0$ with $\alpha$ ranging from 0.3 to 1.8 ($N=10000, L=500$).
   } 
   \label{alpha_001}
\end{figure*}
\begin{figure*}[h]
 \centering
  \subfigure[$\mathrm{MSE}(\alpha)$ for cases of $S(0.8,\beta,1,0)$]{\includegraphics[width=0.35\linewidth]{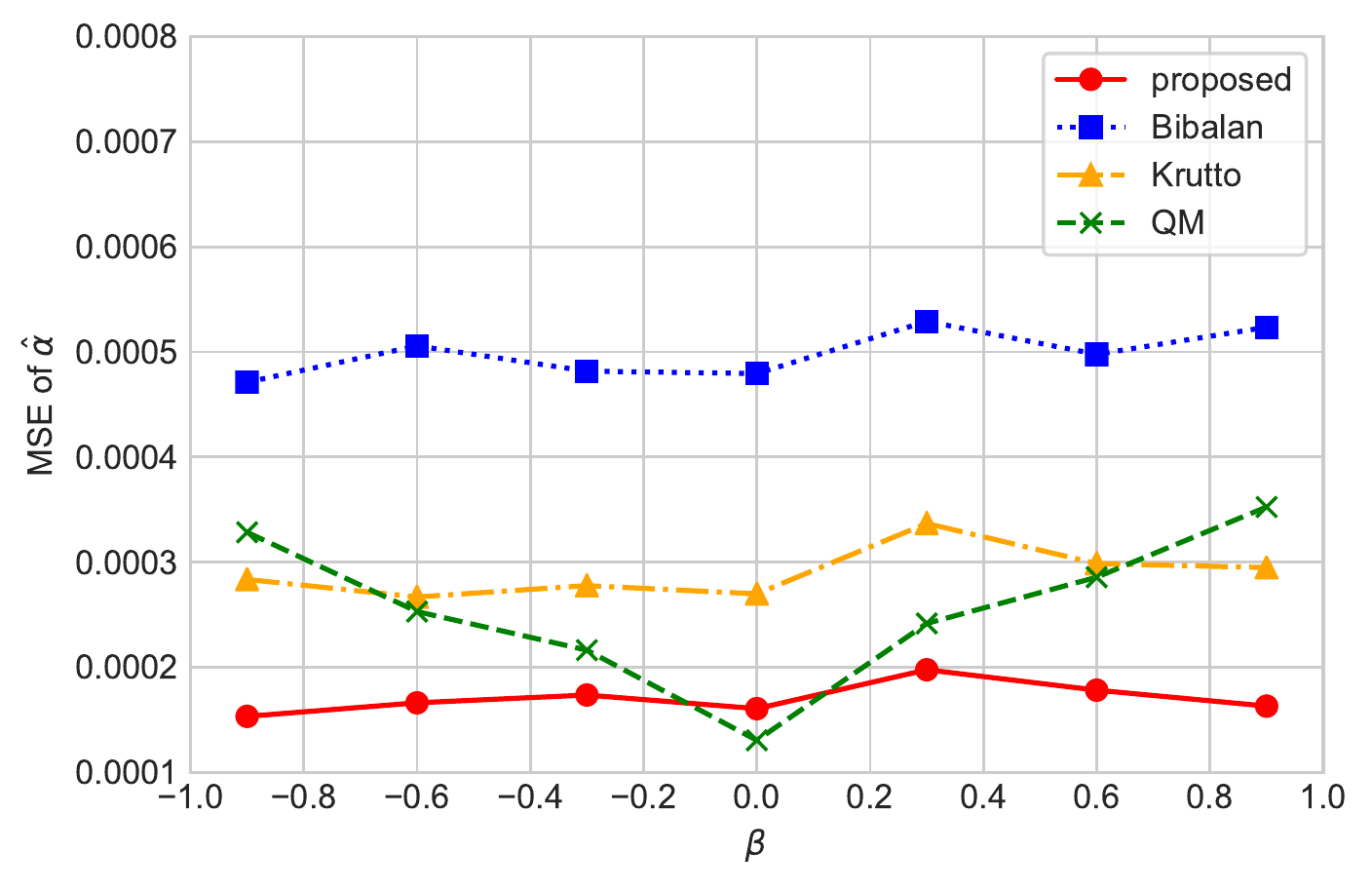}}
  \subfigure[$\mathrm{MSE}(\beta)$ for cases of $S(0.8,\beta,1,0)$]{\includegraphics[width=0.35\linewidth]{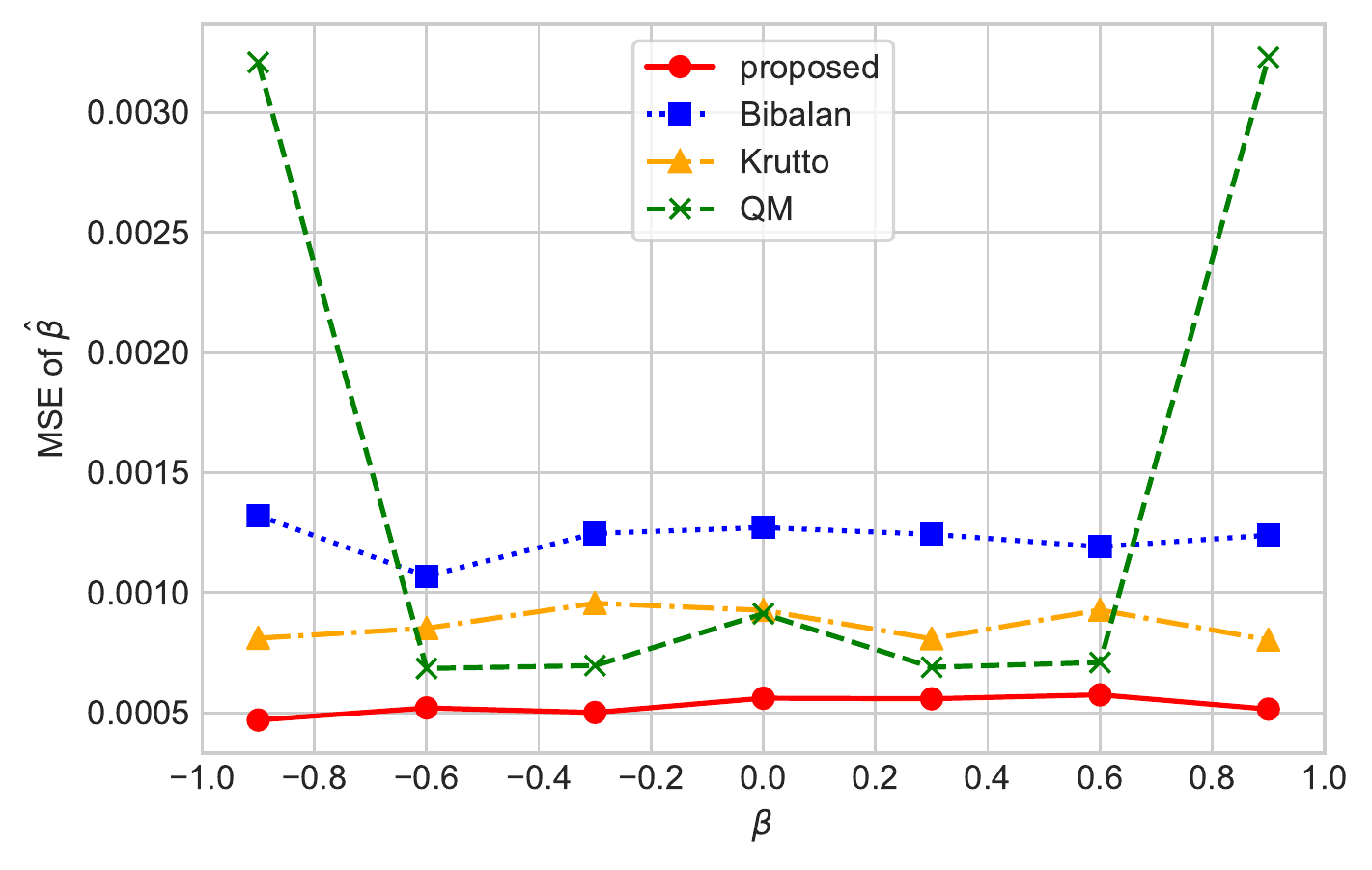}}
  \subfigure[$\mathrm{MSE}(\gamma)$ for cases of $S(0.8,\beta,1,0)$]{\includegraphics[width=0.35\linewidth]{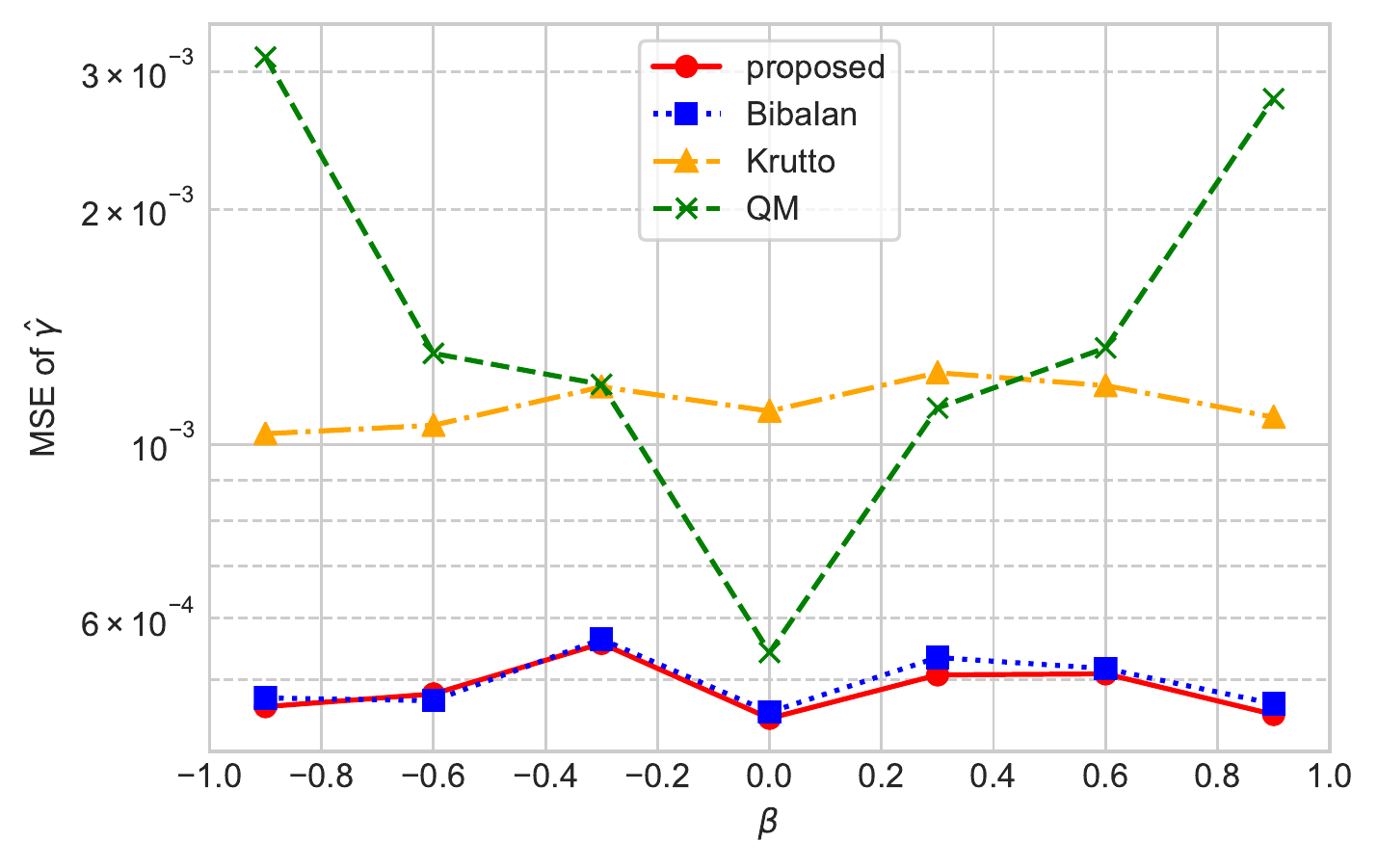}}
  \subfigure[$\mathrm{MSE}(\delta)$ for cases of $S(0.8,\beta,1,0)$]{\includegraphics[width=0.35\linewidth]{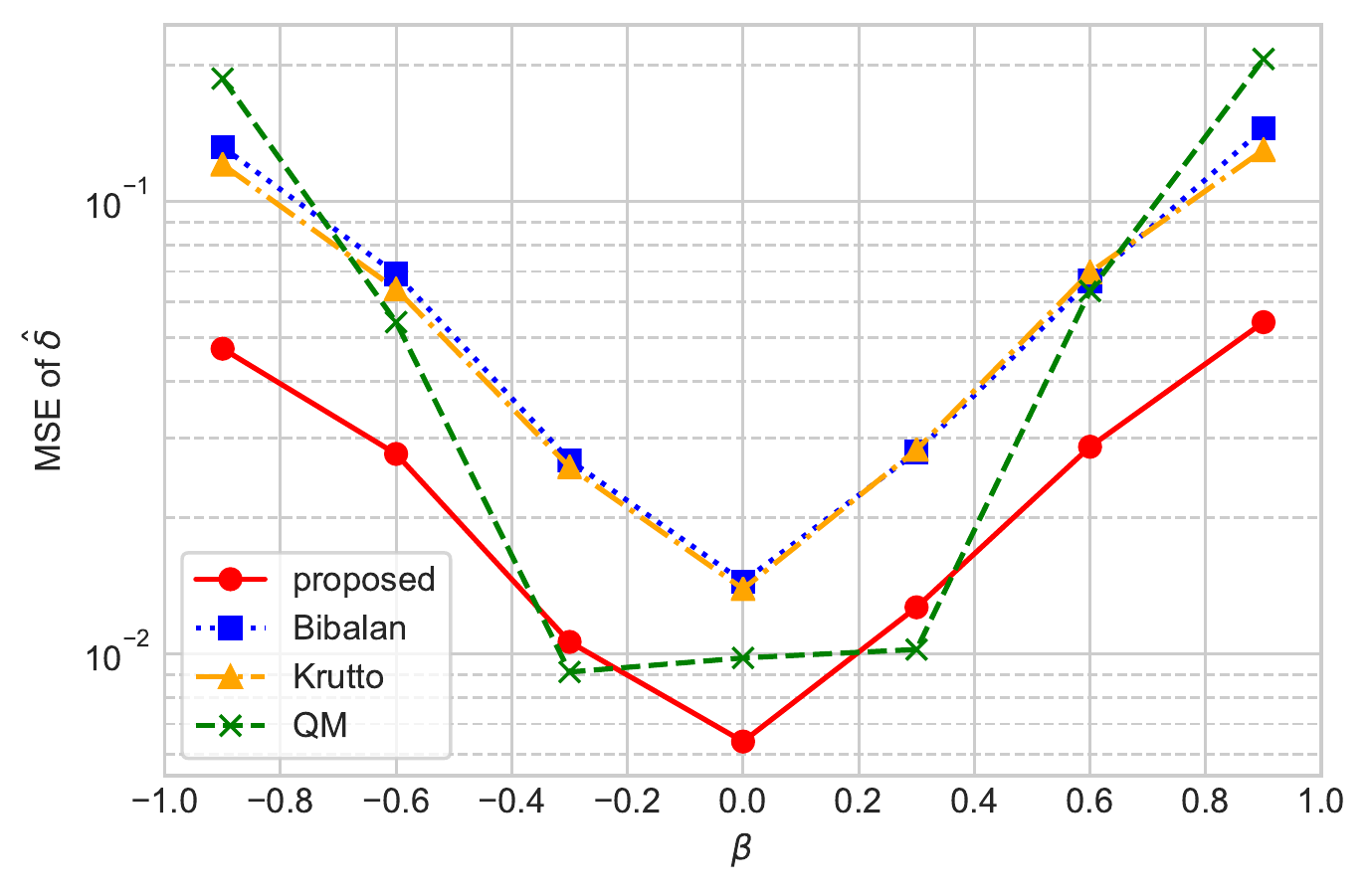}}
  \caption{
  Comparison of the MSEs for the methods based on the proposed approach, Bibalan et al.'s approach, Krutto's approach, and the QM method.
  The MSEs of each stable parameter are studied for cases of parameters $\alpha=0.8, \gamma=1$, and $\delta=0$ with $\beta$ ranging from -0.9 to 0.9 ($N=10000, L=500$).
   } 
   \label{beta_08}
\end{figure*}
\begin{figure*}[h]
 \centering
  \subfigure[$\mathrm{MSE}(\alpha)$ for cases of $S(1.6,\beta,1,0)$]{\includegraphics[width=0.35\linewidth]{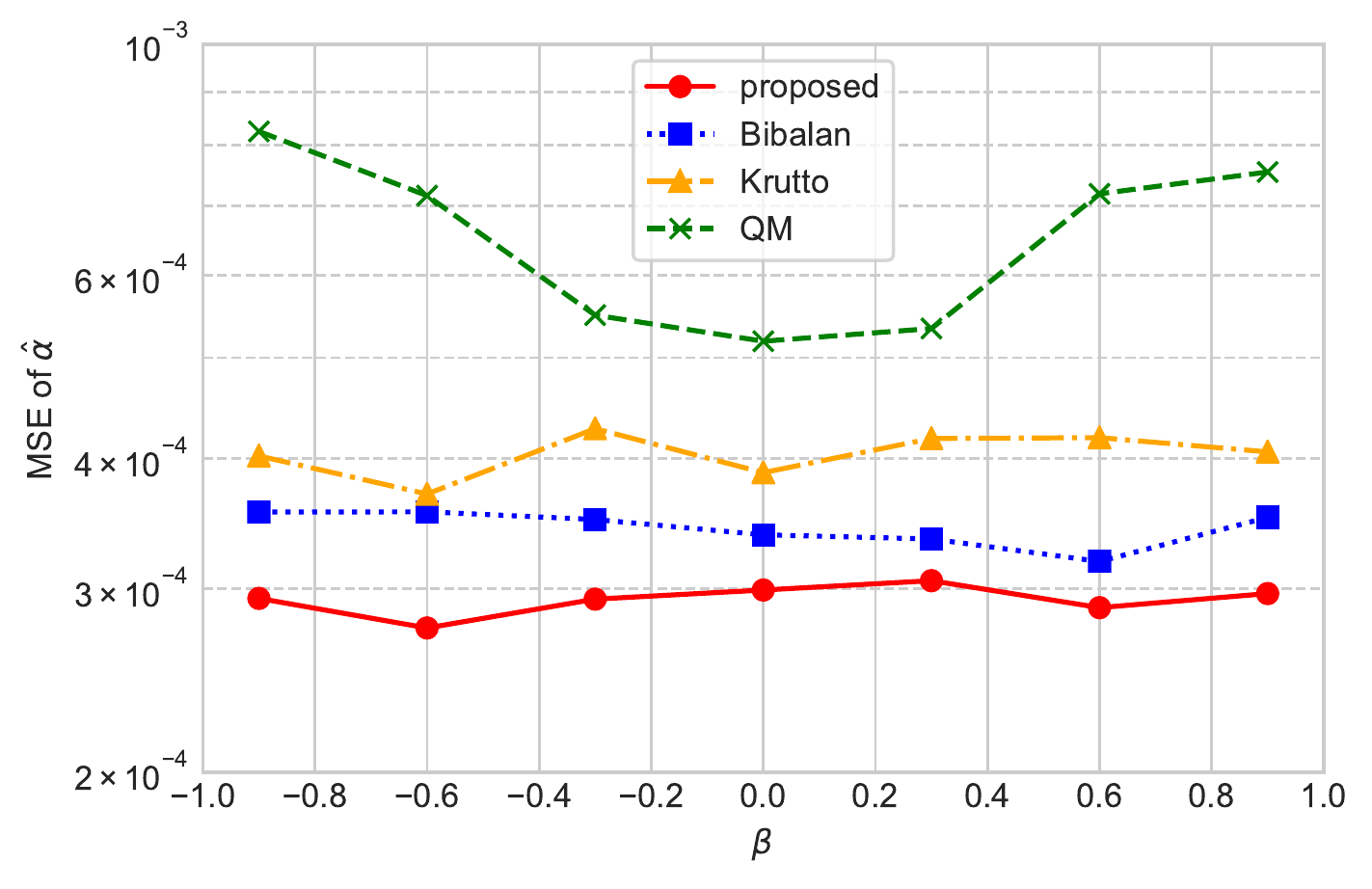}}
  \subfigure[$\mathrm{MSE}(\beta)$ for cases of $S(1.6,\beta,1,0)$]{\includegraphics[width=0.35\linewidth]{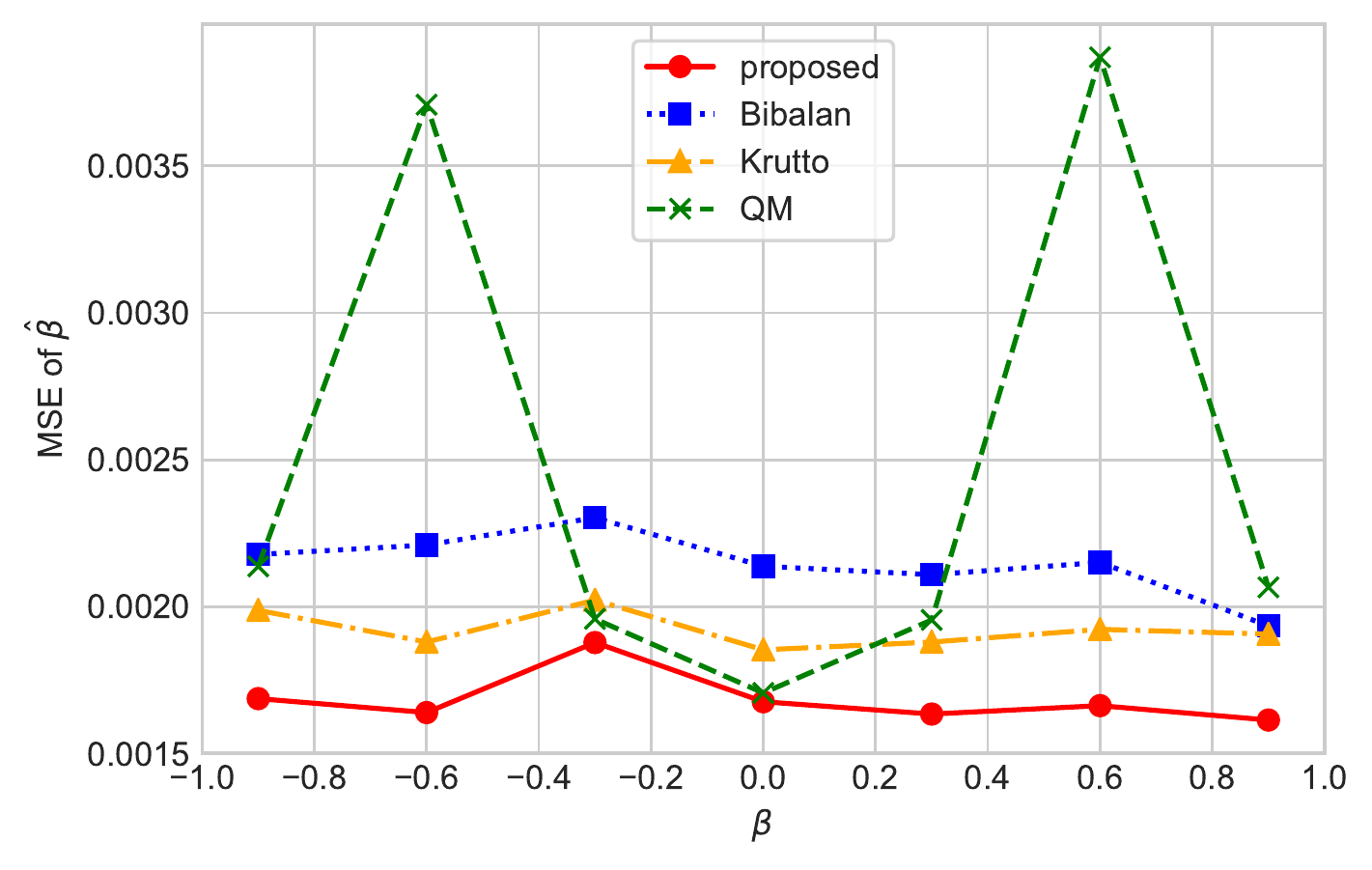}}
  \subfigure[$\mathrm{MSE}(\gamma)$ for cases of $S(1.6,\beta,1,0)$]{\includegraphics[width=0.35\linewidth]{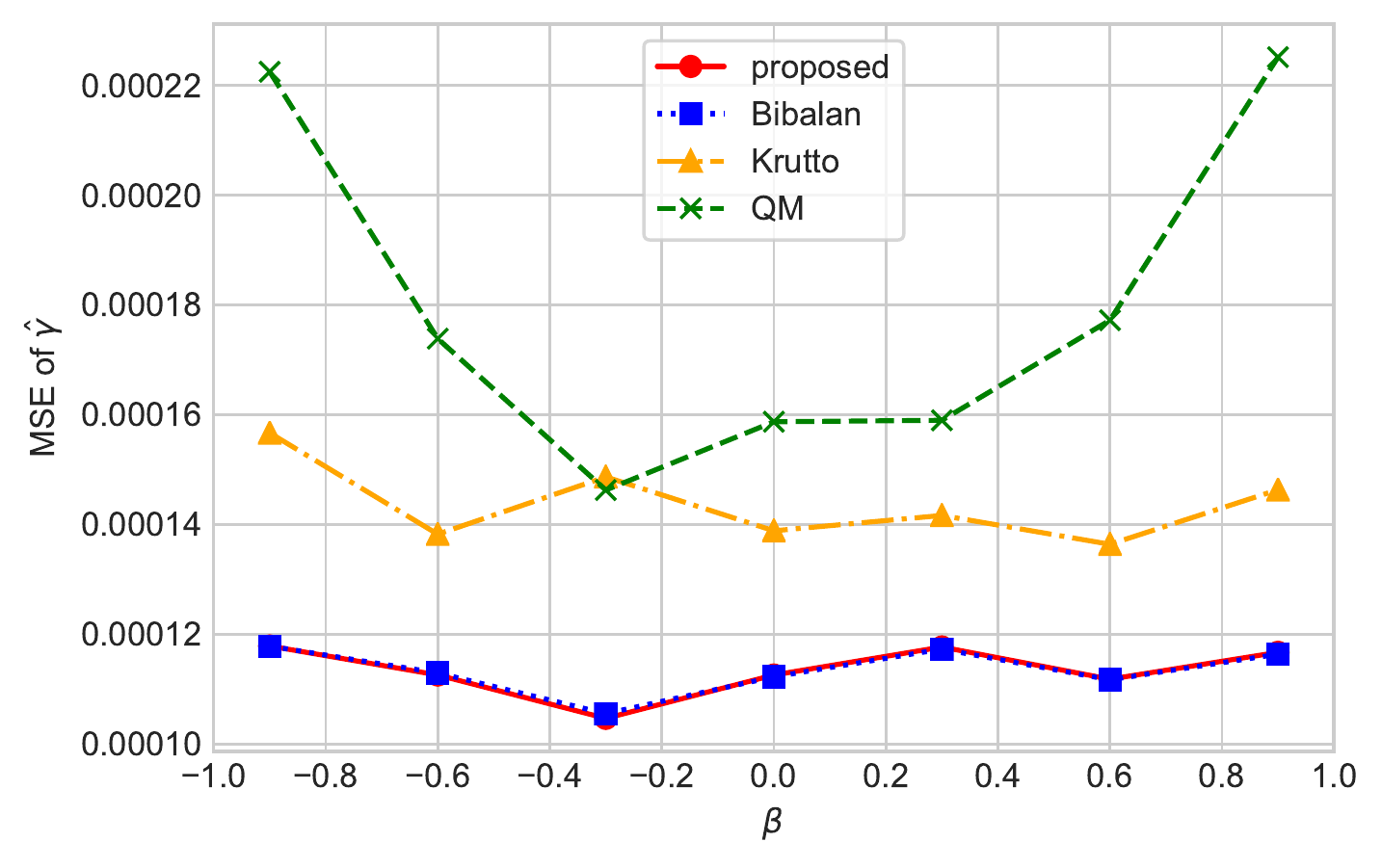}}
  \subfigure[$\mathrm{MSE}(\delta)$ for cases of $S(1.6,\beta,1,0)$]{\includegraphics[width=0.35\linewidth]{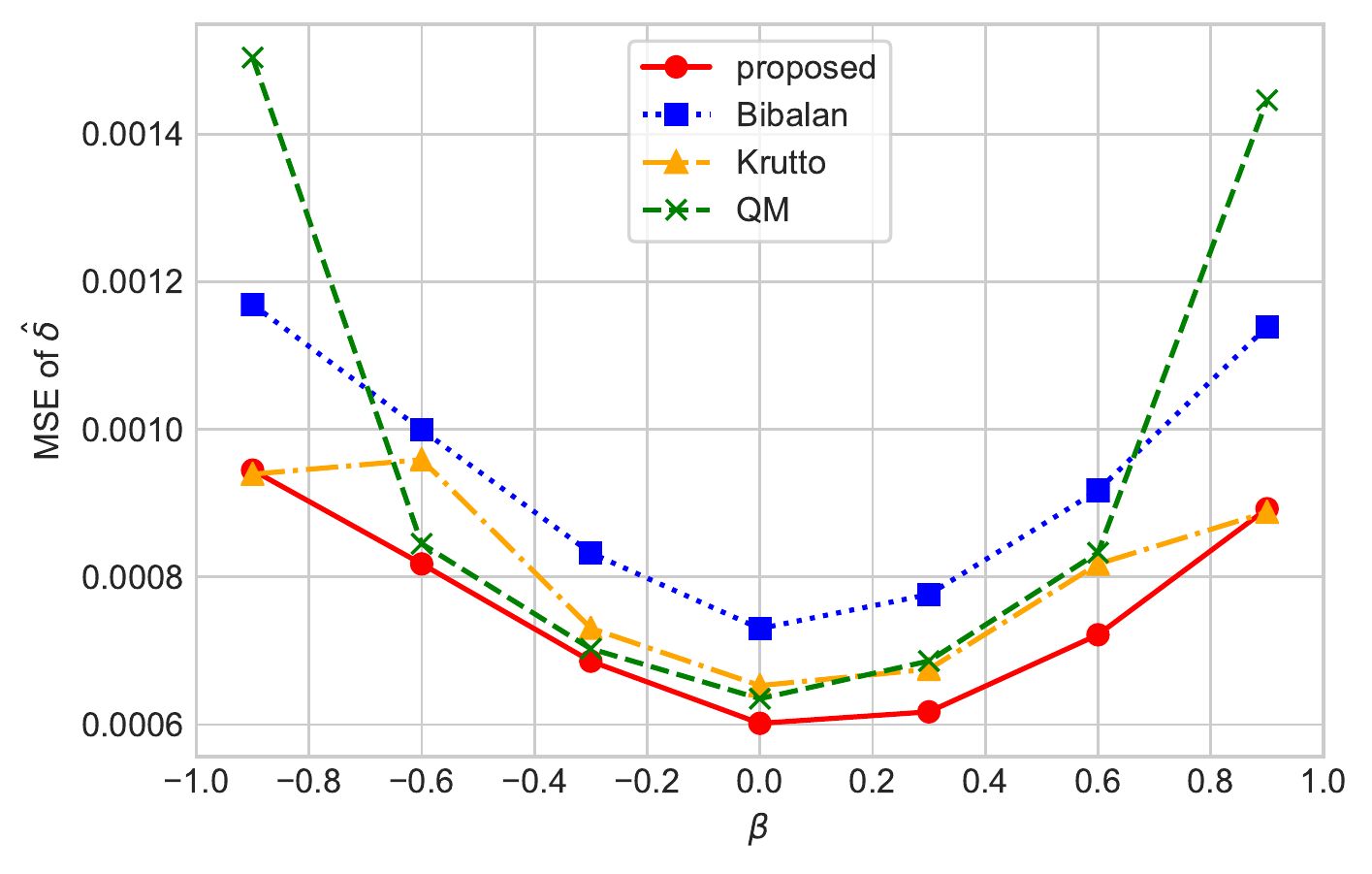}}
  \caption{
  Comparison of the MSEs for the methods based on the proposed approach, Bibalan et al.'s approach, Krutto's approach, and the QM method.
  The MSEs of each stable parameter are studied for cases of parameters $\alpha=1.6, \gamma=1$, and $\delta=0$ with $\beta$ ranging from -0.9 to 0.9 ($N=10000, L=500$).
   } 
   \label{beta_16}
\end{figure*}
%\end{comment}
%
%\clearpage
%%%%%%%%%%%%%%%%%%%%%%%%%%%%%%%%%%%%%%%%%%%%%
% reference
%%%%%%%%%%%%%%%%%%%%%%%%%%%%%%%%%%%%%%%%%%%%%

\end{document}